\documentclass[11pt,a4paper]{article}
\pdfoutput=1

\usepackage{jheppub}
\usepackage[T1]{fontenc}
\usepackage{booktabs}
\usepackage{mathtools}
\usepackage{graphicx}

\newcommand{\eV}{\mathrm{eV}}
\newcommand{\GeV}{\mathrm{GeV}}
\newcommand{\kmsMpc}{\,\mathrm{km}\,\mathrm{s}^{-1}\,\mathrm{Mpc}^{-1}}
\newcommand{\Mpl}{M_{\rm Pl}}
\newcommand{\dd}{\mathrm{d}}
\newcommand{\lcdm}{\Lambda\mathrm{CDM}}

\graphicspath{{figures/}{./}}
\preprint{}

\title{A Multi-Axion Ladder Across Cosmic History\\[0.3em]
{\large From Inflation, BBN, and Early Dark Energy to
Late-Time Accelerated Expansion}}

\author[a,b]{Amartya Sengupta}
\author[a]{Dejan Stojkovic}
\author[c,d]{Adam G. Riess}

\affiliation[a]{Department of Physics, University at Buffalo,
The State University of New York, Buffalo, New York 14260, USA}
\affiliation[b]{Fermi National Accelerator Laboratory,
Batavia, Illinois 60510, USA}
\affiliation[c]{Space Telescope Science Institute,
Baltimore, MD 21218, USA}
\affiliation[d]{Department of Physics and Astronomy,
Johns Hopkins University, Baltimore, MD 21218, USA}

\emailAdd{amartyas@buffalo.edu}
\emailAdd{ds77@buffalo.edu}
\emailAdd{ariess@stsci.edu}

\abstract{
What if dark energy is recurrent throughout cosmic history? In this picture, episodes of scalar-field dark energy become
less surprising and more natural features of cosmic evolution. We construct a homogeneous multi-axion cosmology with a transient
contribution during Big Bang nucleosynthesis, two early dark energy components before recombination, and a thawing field
that supplies the present dark-energy density. As Hubble friction weakens, the fields begin to roll at successive epochs set by
their potential-curvature scales. Their initial displacements affect the rolling delays and peak energy fractions. The transient
fields have third-power cosine potentials, whose sextic minima allow faster-than-radiation dilution during rapid, small-amplitude
oscillations. All four fields are evolved in a common Friedmann background, with their first roll and subsequent dynamics resolved numerically. With reference matter and radiation densities taken from Planck 2018, the benchmark has a
nucleosynthesis-era peak fraction of approximately \(0.99\%\) near \(z=10^9\). The two early dark energy fields peak near
\(z=7.9\times10^3\) and \(2.4\times10^3\), with individual fractions of \(8.6\%\) and \(7.8\%\); their combined fraction
reaches \(9.7\%\). The late-time field is normalized to supply a present fraction of approximately \(0.685\), while the three
transients leave a combined fraction of approximately \(6.0\times10^{-7}\). An aligned two-axion example illustrates
the enhanced field range available for an inflationary extension. We also outline how searches for transient contributions to
the expansion rate at other epochs could constrain additional axion scales.
}


\begin{document}
\maketitle

% ============================================================
\section{Introduction}
\label{sec:intro}
% ============================================================

Cosmological observations constrain the expansion history
through physical processes operating at widely separated
epochs. Spatially flat \(\lcdm\), with a cosmological constant
and cold dark matter, remains the reference model. The Hubble
tension concerns the difference between the present expansion
rate inferred from Planck cosmic microwave background (CMB)
data in this model and the value measured with the local
distance ladder
\cite{Planck:2018vyg,Riess:2021jrx,H0DN:2025lyy,
DiValentino:2021izs,Knox:2019rjx}. Big Bang nucleosynthesis
(BBN) probes much earlier expansion through the primordial
light-element abundances
\cite{Cooke:2017cwo,Pitrou:2018cgg,Iocco:2008va}.
At late times, baryon acoustic oscillation (BAO) measurements
from the Dark Energy Spectroscopic Instrument (DESI) constrain
cosmological distances and the expansion rate. Together with
CMB observations and Type Ia supernova distances, these
measurements can favor an evolving dark-energy equation of
state, although the strength of this preference varies across
different analyses
\cite{DESI:2024mwx,DESI:2025zgx,DESI:2025fii,Brout:2022vxf}.
These observations motivate examining whether scalar fields
can affect more than one stage of cosmic expansion through
a common particle-physics mechanism.

During BBN, the competition between reaction rates and
expansion determines neutron survival and the extent of
nuclear processing. Before recombination, the available
propagation time sets the distance traveled by acoustic waves
through the photon--baryon plasma. Late-time distances
integrate the expansion history, while structure growth also
depends on the evolving matter fraction and gravitational
potentials. Consequently, two transients with the same peak
energy fraction can have different observational consequences
if they occur at different times or dilute at different rates.

A scalar field initially displaced from its potential minimum
can produce a transient contribution to the total energy density.
While Hubble damping keeps the field nearly frozen, its energy
density remains approximately constant and potential dominated.
Its fractional contribution therefore grows as matter and
radiation dilute with expansion. As the field rolls down its potential, part of its potential
energy is converted into kinetic energy. The subsequent dilution depends on the potential near its minimum and can reduce the scalar fraction again without a decay into other particles.During each transient, the scalar energy fraction grows before
the subsequent dilution reduces it relative to the background.
The rolling dynamics and later evolution determine the peak
fraction, the duration of the contribution, and the residual
density.

Early dark energy (EDE) has been studied extensively as an
application of this mechanism before recombination.
A temporary scalar contribution increases the expansion rate
and can shorten the sound horizon, allowing a larger
CMB-inferred Hubble constant in some fits
\cite{Karwal:2016vyq,Poulin:2018cxd,Smith:2019ihp,
Poulin:2023lkg}. The acoustic peaks, photon diffusion,
gravitational-potential evolution, and matter clustering
provide complementary constraints on this possibility
\cite{Hill:2020osr,Murgia:2020ryi,Ivanov:2020ril,
DAmico:2020ods,Smith:2020rxx,Herold:2021ksg,
Simon:2022adh,McDonough:2023qcu}. Recent analyses continue
to find sensitivity to the observations included
\cite{Poulin:2025nfb,SPT-3G:2025vyw}.
The time dependence of the contribution is particularly
relevant in multi-field models: a recent study finds that
a second axion improves the high-multipole CMB fit relative
to the single-field model considered there
\cite{Bella:2026zuk}. This result motivates examining how
the separation and overlap of scalar contributions shape
the pre-recombination expansion history.

Connections between early- and late-time evolution have
also been explored through interactions within the dark
sector
\cite{Khoury:2026svx,SevillanoMunoz:2026jgk},
including a dark axion coupled to dark baryons
\cite{Khoury:2025txd}. A spectrum of axion-like fields
offers another physical connection between epochs.
Different members of the spectrum can contribute at
different stages of expansion, with their evolution linked
by the total energy density that determines the Hubble rate.

The particle-physics motivation for such a spectrum comes
from approximate shift symmetries and non-perturbative
symmetry breaking. Weak explicit breaking allows small
potential-curvature scales to remain stable against
radiative corrections. String compactifications can contain
many axions, whose non-perturbative potential amplitudes
depend exponentially on the underlying instanton actions.
Differences in these actions can generate widely separated
curvature scales even when the decay constants are comparable
\cite{Svrcek:2006yi,Arvanitaki:2009fg,Marsh:2015xka,
Li:2025cep}. Each scale corresponds to a characteristic
response time for a displaced field. As the expansion slows,
fields with progressively longer response times can develop
appreciable motion. This connection between a hierarchy of
axion scales and successive cosmological epochs motivates
the mass-ladder picture. The initial displacement and the
shape of the potential then determine how much energy
participates in each episode and how long its influence lasts.

Kamionkowski, Pradler, and Walker developed an axiverse
model in which successive fields become dynamical as the
Hubble rate decreases
\cite{Kamionkowski:2014zda}. They addressed the late onset
of cosmic acceleration statistically, using a hierarchy
of axion scales and randomly distributed initial
misalignment angles. They also considered potentials
proportional to \([1-\cos(\phi/f)]^3\) to suppress the
relic densities of fields active at earlier times.
Near their minima, these potentials are sextic.
Rapid, small-amplitude oscillations then have an averaged
equation of state \(\langle w\rangle=1/2\), with
\(\rho_\phi\propto a^{-9/2}\). This dilution is faster
than that of radiation and allows an earlier contribution
to become small through its scalar dynamics alone.
The mechanism provides a concrete basis for studying
several dark-energy contributions within one expansion
history.

We develop a four-field example that connects the BBN era,
the interval around matter--radiation equality, and
late-time acceleration. The three earlier fields use the
third-power cosine potential, while an ordinary cosine
supports the late-time thawing component. With the potential
parameters and initial angles specified, we follow the
energy transfer between potential and kinetic terms
throughout the post-inflationary evolution. Each field
contributes to the expansion rate that damps the motion
of the others; this gravitational coupling is especially
relevant while the two EDE components overlap.
The calculation relates their peak amplitudes and timing
to the densities that survive through the subsequent
radiation, matter, and dark-energy eras.

For the benchmark, the BBN-era contribution reaches about
one per cent of the total density. The two EDE fields attain
their individual maxima on opposite sides of
matter--radiation equality, and their overlap produces
a combined maximum of approximately \(9.7\%\).
The late-time normalization sets the present dark-energy
fraction to approximately \(0.685\), whereas the three
earlier fields together account for only
\(6.0\times10^{-7}\) of the density today. A scan of the initial hilltop offsets further
examines how the transient amplitudes and epochs depend
on the adopted displacements.

The hierarchy also suggests an empirical question:
which other epochs could accommodate an appreciable
axion contribution? Constraints on the amplitude and
duration of localized excess expansion would identify
where additional fields remain cosmologically accessible.
For a specified potential, the relation between an
observed epoch and an axion curvature scale depends on
the delay between the weakening of Hubble damping and
the maximum density. We discuss how this connection
could extend the study from the selected benchmark
epochs to a search for additional axion contributions at other
epochs of cosmic history.

Axion alignment motivates an inflationary extension of the
framework by enlarging the effective field range along a
light axion field direction. When two periodic potentials
depend on nearly parallel combinations of canonical
fields, a light axion field direction can acquire a super-Planckian
effective decay constant even though the microscopic
decay constants are sub-Planckian
\cite{Kim:2004rp,Choi:2014rja,Long:2018nsl}.
The enhanced field range motivates the inflaton candidate
discussed in section~\ref{subsec:inflation_connection}.
Appendix~\ref{app:inflation} develops the alignment example
and examines the inflationary-scale and reheating estimates
relevant to the later fields.

\begin{figure}[htbp]
\centering
\includegraphics[width=\textwidth]{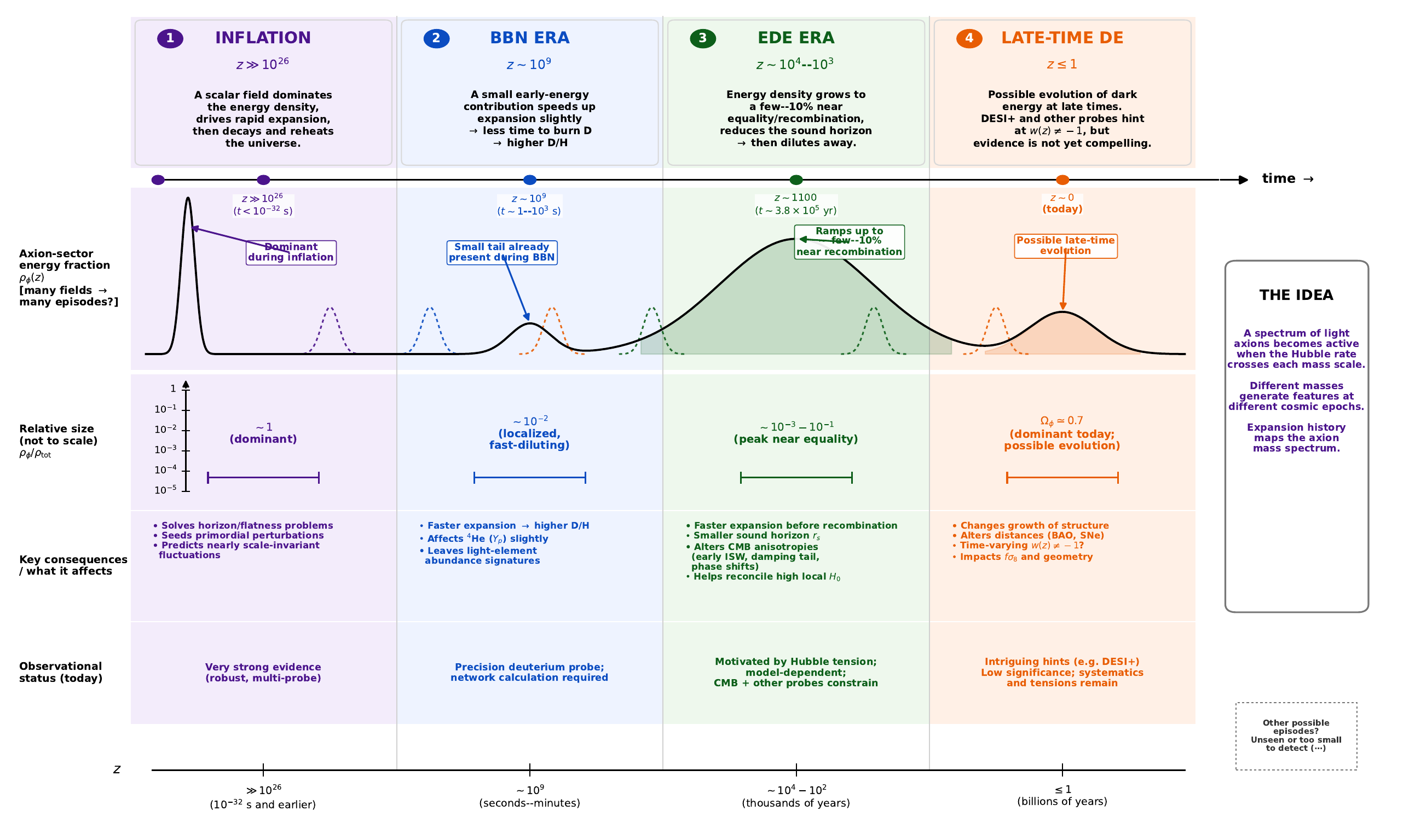}
\caption{Qualitative arrangement of the axion contributions
across cosmic history, including an inflationary
sector and the BBN, pre-recombination EDE, and late-time
dark-energy components. For the post-inflationary fields,
the curvature hierarchy sets the characteristic ordering
of epochs, while the initial displacements affect the
rolling delays and peak fractions. The potential near
each minimum governs the subsequent dilution.
The schematic is not to scale; the numerical evolution
is presented in table~\ref{tab:benchmark} and
figure~\ref{fig:omega_ladder}.}
\label{fig:axion_ladder_schematic}
\end{figure}

We introduce the multi-axion model in section~\ref{sec:model}
and examine the onset of rolling and subsequent dilution
in section~\ref{sec:dynamics}.
Section~\ref{sec:numerical_benchmark} presents the benchmark
evolution. Its implications for BBN, EDE, and late-time
acceleration are discussed in section~\ref{sec:epochs},
followed by the particle-physics considerations and
initial conditions in section~\ref{sec:consistency}.
Section~\ref{sec:predictions} examines the cosmological
tests and the search for additional rungs, and
section~\ref{sec:discussion} brings together the results
and their implications. The appendices contain the
hilltop sensitivity scan, the aligned inflationary
example, and the numerical equations and convergence
checks.

% ============================================================
\section{Multi-axion model and characteristic scales}
\label{sec:model}
% ============================================================

\subsection{Post-inflationary fields and background equations}

We consider four canonical axion-like pseudoscalars
\(\phi_i\), with \(i\in\{\mathrm{BBN},\mathrm{EDE1},
\mathrm{EDE2},\mathrm{DE}\}\); the final label denotes the
late-time dark-energy field. We use units \(c=\hbar=1\) and
metric signature \((-+++)\). The post-inflationary action is
\begin{equation}
S_{\rm post}=\int\dd^4x\,\sqrt{-g}
\left[\frac{\Mpl^2}{2}R
-\frac12\sum_i g^{\mu\nu}\partial_\mu\phi_i\partial_\nu\phi_i
-\sum_i V_i(\phi_i)\right]+S_{m+r},
\label{eq:action}
\end{equation}
where \(g\) is the metric determinant, \(R\) is the Ricci scalar,
\(S_{m+r}\) contains matter and radiation, and
\(\Mpl=(8\pi G)^{-1/2}=2.435\times10^{18}\,\GeV\) is the
reduced Planck mass. We assume
\begin{equation}
V_i(\phi_i)=\Lambda_i^4[1-\cos\Theta_i]^{n_i},
\qquad
\Theta_i(t)\equiv\frac{\phi_i(t)}{f_i},
\qquad
\theta_i\equiv\Theta_i(t_{\rm ini}).
\label{eq:axion_potential}
\end{equation}
The decay constant \(f_i\) sets the period \(2\pi f_i\),
\(\Lambda_i\) sets the potential scale, and the positive integer \(n_i\) fixes the leading power near the minimum. The angle \(\theta_i\) specifies the initial displacement at the starting time \(t_{\rm ini}\). The potential height is
\(V_i(\pi f_i)=2^{n_i}\Lambda_i^4\). We choose \(n_i=3\) for the three transients and \(n_{\rm DE}=1\).

Non-perturbative effects can generate a broad hierarchy of axion potential scales \cite{Svrcek:2006yi,Arvanitaki:2009fg,Marsh:2015xka}.
For our benchmark, we take these scales to be temperature
independent and neglect kinetic and potential mixing between
the fields.\footnote{A related issue arises in non-Abelian
flavour models, where mixed scalar operators can break the
residual symmetries of the separate scalar potentials and
reorient their vacuum alignments
\cite{deMedeirosVarzielas:2025byb}}. This assumption must hold throughout their motion, not just near a potential minimum. In particular, the transient potentials have vanishing curvature at their sextic minima, so the vacuum mass matrix alone cannot constrain higher-order interactions between them. Each field contributes to the Hubble rate through the Friedmann equation and thereby affects the Hubble friction acting on the other fields.

The expansion
\begin{equation}
1-\cos\Theta_i=\frac{\Theta_i^2}{2}
-\frac{\Theta_i^4}{24}+\mathcal O(\Theta_i^6)
\label{eq:cosine_expansion}
\end{equation}
shows the distinction between the late-time and transient fields. Near its minimum, \(|\Theta_{\rm DE}|\ll1\), the ordinary cosine potential is quadratic to leading order:
\begin{equation}
V_{\rm DE}=\frac12m_{\rm DE}^2\phi_{\rm DE}^2
+\mathcal O(\Lambda_{\rm DE}^4\Theta_{\rm DE}^4),
\qquad
m_{\rm DE}=\frac{\Lambda_{\rm DE}^2}{f_{\rm DE}}.
\label{eq:n1_mass}
\end{equation}
For the transient fields, the expansion around
\(\Theta_i=0\) instead begins at sextic order:
\begin{equation}
V_i=\frac{\Lambda_i^4}{8}\Theta_i^6
-\frac{\Lambda_i^4}{32}\Theta_i^8
+\mathcal O(\Lambda_i^4\Theta_i^{10}),
\qquad
V_{i,\phi\phi}(0)=0,
\qquad n_i=3.
\label{eq:sextic_minimum}
\end{equation}
The first potential has a vacuum particle mass; the second has an amplitude-dependent oscillation frequency instead.

The derivatives of the periodic potential are
\begin{align}
V_{i,\phi}&=\frac{n_i\Lambda_i^4}{f_i}
(1-\cos\Theta_i)^{n_i-1}\sin\Theta_i,
\nonumber\\
V_{i,\phi\phi}(\pi f_i)&
=-n_i2^{n_i-1}\frac{\Lambda_i^4}{f_i^2}.
\label{eq:force_and_hilltop_curvature}
\end{align}
The negative curvature drives the hilltop instability. For the three transients we define
\begin{equation}
m_{{\rm eff},i}^2\equiv
\left|V_{i,\phi\phi}(\pi f_i)\right|
=12\frac{\Lambda_i^4}{f_i^2}.
\label{eq:n3_meff}
\end{equation}
This fixed scale differs from the instantaneous curvature along the trajectory,
\begin{equation}
V_{i,\phi\phi}(\phi_i(t))
=\frac{3\Lambda_i^4}{f_i^2}
(1-\cos\Theta_i)^2(2+3\cos\Theta_i),
\qquad n_i=3,
\label{eq:transient_curvature}
\end{equation}
which changes sign during the roll and vanishes at the minimum. The transient scales \(m_{{\rm eff},i}\) characterize the hilltop curvatures; the equations of motion retain the full periodic potentials. For the late-time field, \(m_{\rm DE}\) is the vacuum mass.

We normalize the scale factor by \(a(t_0)=1\), define redshift through \(1+z=a^{-1}\), and write the Hubble rate as
\(H=\dot a/a\). A subscript \(0\) denotes the present epoch. In a spatially flat Friedmann--Lema\^{\i}tre--Robertson--Walker background, the scalar energy densities, pressures, and equations of state are
\begin{equation}
\rho_{\phi_i}=\frac12\dot\phi_i^2+V_i,
\qquad P_{\phi_i}=\frac12\dot\phi_i^2-V_i,
\qquad w_i=\frac{P_{\phi_i}}{\rho_{\phi_i}}.
\label{eq:rho_pressure}
\end{equation}
The homogeneous evolution is governed by the coupled Friedmann, Klein--Gordon, and Raychaudhuri equations,
\begin{align}
H^2
&=\frac{1}{3\Mpl^2}
\left(\rho_r+\rho_m+\sum_i\rho_{\phi_i}\right),
\label{eq:friedmann}\\
\ddot\phi_i+3H\dot\phi_i+V_{i,\phi}
&=0,
\label{eq:kg}\\
\dot H
&=-\frac{1}{2\Mpl^2}
\left(\frac{4}{3}\rho_r+\rho_m+\sum_i\dot\phi_i^2\right).
\label{eq:raychaudhuri}
\end{align}
Here \(\rho_r\) and \(\rho_m\) denote radiation and
non-relativistic matter. Their sum with the scalar densities is
\(\rho_{\rm tot}\), while
\(P_{\rm tot}=\rho_r/3+\sum_iP_{\phi_i}\).
Multiplying eq.~\eqref{eq:kg} by \(\dot\phi_i\) we get
\begin{equation}
\dot\rho_{\phi_i}=-3H\dot\phi_i^2
=-3H(1+w_i)\rho_{\phi_i}.
\label{eq:scalar_conservation}
\end{equation}
Thus the scalar energy density decreases or stays constant
during expansion; the transient enhancement concerns its
fraction of the total density. We use
\begin{equation}
\Omega_i(z)\equiv\frac{\rho_{\phi_i}}{3\Mpl^2H^2},
\qquad \Omega_{r,m}(z)\equiv\frac{\rho_{r,m}}{3\Mpl^2H^2},
\label{eq:fractions}
\end{equation}
so flatness implies
\(\Omega_r+\Omega_m+\sum_i\Omega_i=1\).

We adopt the reference values
\begin{equation}
H_0=67.4\kmsMpc,\quad
\Omega_{m0}=0.315,\quad
\Omega_{r0}h^2=4.18\times10^{-5},\quad
h\equiv\frac{H_0}{100\kmsMpc},
\label{eq:reference_parameters}
\end{equation}
using the Planck 2018 base-\(\lcdm\) parameters as inputs
\cite{Planck:2018vyg}. The matter and radiation densities evolve as
\begin{equation}
\rho_m=3\Mpl^2H_0^2\Omega_{m0}a^{-3},
\qquad
\rho_r=3\Mpl^2H_0^2\Omega_{r0}a^{-4}.
\label{eq:reference_fluids}
\end{equation}
The fixed \(a^{-4}\) radiation law defines the background
approximation used here. A more detailed evolution must account for electron--positron annihilation, neutrino decoupling, and the later non-relativistic transition of massive neutrinos. The BBN abundance calculation also requires a consistent relation between photon temperature and scale factor.

During radiation domination, for small scalar fractions,
\begin{equation}
H(z)\simeq H_0\sqrt{\Omega_{r0}}(1+z)^2,
\qquad H_0\sqrt{\Omega_{r0}}=1.38\times10^{-35}\,\eV.
\label{eq:rad_H}
\end{equation}
A nominal curvature-crossing redshift is defined by
\begin{equation}
H(z_{\times,i})=m_{{\rm eff},i}.
\label{eq:clock_model}
\end{equation}
The radiation-era estimate is therefore
\begin{equation}
1+z_{\times,i}\simeq
\left(\frac{m_{{\rm eff},i}}
{1.38\times10^{-35}\,\eV}\right)^{1/2}.
\label{eq:zstar}
\end{equation}
Near matter--radiation equality, matter and radiation have
comparable densities, while the EDE fields provide an additional contribution to the Hubble rate. An appreciable frozen field increases Hubble damping and changes
the subsequent roll. We include this backreaction in the coupled evolution and identify the fractional-density maxima using the peak condition derived in section~\ref{sec:dynamics}.

In coupled EDE models, the timing can be linked to
neutrinos becoming non-relativistic \cite{Sakstein:2019fmf}, or to the evolving dark-matter density \cite{Karwal:2021vpk}.
For the potentials considered here, the characteristic
epochs are set by their curvature scales and initial
displacements, with the shared expansion affecting
the subsequent rolling.

\subsection{An aligned inflationary sector}
\label{subsec:inflation_connection}

An aligned two-axion sector can provide an inflationary extension of the ladder. Natural inflation associates the inflaton with a periodic field direction \cite{Freese:1990rb}. For two axions, 
alignment can enlarge the effective period
when the two potential terms depend on nearly parallel
combinations of the fields \cite{Kim:2004rp,Choi:2014rja}.
If one combination remains confined near its minimum,
motion along the remaining direction changes the phase
of the second potential only slowly, producing an extended
field range. In the aligned example discussed in appendix~\ref{app:inflation}, both microscopic decay constants are \(0.5\Mpl\), while the effective decay constant along the light direction is \(f_{\rm eff}\simeq7.43\Mpl\). The field along this slowly
varying direction is the candidate inflaton, with an
effective period of \(2\pi f_{\rm eff}\).

The connection to the four later fields follows from their much smaller curvature scales. For the assumed potentials, Hubble friction strongly suppresses their motion during inflation and at the fiducial reheating temperature. Their displacements can then supply the initial potential energies of the BBN, EDE, and late-time components, provided thermal and nonthermal interactions preserve the field values and potentials through reheating. As the expansion rate subsequently decreases, the fields roll according to their curvature scales and initial misalignments.

The aligned sector illustrates the enhancement of the axion
field range, while the numerical solution follows the four
post-inflationary fields after reheating. We adopt \(r=0.01\)
as a separate fiducial input for the inflationary normalization and the initial fluctuations discussed in
section~\ref{subsec:isocurvature}. Appendix~\ref{app:inflation} derives the enhanced period, evaluates the corresponding inflationary scales, and examines the Hubble damping of the four later fields at the adopted reheating temperature.

% ============================================================
\section{Hilltop evolution and transient dilution}
\label{sec:dynamics}
% ============================================================

For the transient fields, we define the time-dependent displacement
\(x_i(t)=\pi-\Theta_i(t)\) and its initial value
\(\delta_i=\pi-\theta_i\). Near the maximum,
\begin{equation}
V_i=2^{n_i}\Lambda_i^4
\left[1-\frac{n_i}{4}x_i^2+\mathcal O(x_i^4)\right].
\label{eq:hilltop_expansion}
\end{equation}
The linearized equations for the transient fields take the form
\begin{equation}
\ddot x_i+3H\dot x_i-m_{{\rm eff},i}^2x_i\simeq0.
\label{eq:hilltop_linearized}
\end{equation}
The negative curvature permits growth before the nominal crossing, although Hubble friction slows the displacement when \(H\gg m_{{\rm eff},i}\). A smaller initial offset needs more amplification before the roll becomes nonlinear. During this delay,
\begin{equation}
\rho_{\phi_i}\simeq\Lambda_i^4(1-\cos\theta_i)^3,
\qquad w_i\simeq-1.
\label{eq:frozen_density}
\end{equation}
As matter and radiation dilute, a nearly frozen scalar field
contributes an increasing fraction of the total energy density. At the exact potential maximum, the force vanishes, so a field with zero initial velocity remains there in the homogeneous equations. A finite initial offset allows the displacement to grow and the field to roll. While the field remains frozen and subdominant, its fraction grows approximately as \(a^4\) during radiation domination and as \(a^3\) during matter domination. Delaying the roll allows the background to dilute further while the scalar density changes little, so even a modest delay can substantially increase the peak fraction.

For a field still approximately frozen at its nominal crossing,
\begin{align}
\Omega_{i,\times}&\simeq
\frac{\Lambda_i^4(1-\cos\theta_i)^3}
{3\Mpl^2m_{{\rm eff},i}^2}
=\frac{f_i^2}{36\Mpl^2}(1-\cos\theta_i)^3,
\nonumber\\
\Omega_{i,\times}&\longrightarrow\frac{2f_i^2}{9\Mpl^2}
\quad(\theta_i\longrightarrow\pi).
\label{eq:crossing_fraction}
\end{align}
Because the potential scale sets both the frozen energy density and the hilltop curvature, it cancels from this estimate, which at the benchmark offset is \(3.32\times10^{-4}\) for BBN and \(3.40\times10^{-3}\) for either EDE field, well below the calculated peaks. The difference reflects the rolling delay, during which the background dilutes while the scalar density remains nearly constant, allowing its fractional contribution
to grow before reaching a later, higher maximum.

The evolution of the scalar fraction follows from
eq.~\eqref{eq:scalar_conservation} and total energy conservation:
\begin{equation}
\frac{\dd\Omega_i}{\dd\ln a}
=3\Omega_i(w_{\rm tot}-w_i),
\qquad w_{\rm tot}\equiv\frac{P_{\rm tot}}{\rho_{\rm tot}}.
\label{eq:fraction_peak_condition}
\end{equation}
We denote the maximum fraction by \(\Omega_i^{\rm peak}\)
and its redshift by \(z_{{\rm peak},i}\). A nonzero interior
maximum satisfies \(w_i=w_{\rm tot}\), with the derivative
changing from positive to negative. The field fraction grows
while its equation of state lies below \(w_{\rm tot}\), and
falls once the kinetic energy raises \(w_i\) above
\(w_{\rm tot}\). We use this condition to identify
fractional-density maxima along the scalar trajectories.

In the limiting case of a subdominant field in a radiation or
matter background, the peak condition corresponds respectively to
\begin{equation}
\frac{\dot\phi_i^2}{2V_i}\simeq2
\quad\text{or}\quad
\frac{\dot\phi_i^2}{2V_i}\simeq1.
\label{eq:peak_kinetic_ratio}
\end{equation}
The kinetic energy is therefore already appreciable at the
maximum; the field is no longer frozen. Near equality, and when the EDE fields overlap, neither limiting ratio is exact.
Their shared contribution changes \(w_{\rm tot}\), so the same initial potential can reach a different peak in a different background.

Near the minimum, rapid motion allows a useful analytic description. During one oscillation, neglecting the change of amplitude and scale factor, the cycle average of
\(\dd(\phi_i\dot\phi_i)/\dd t\) vanishes. In the rapid-oscillation regime, the averaged field dynamics obey the virial relation
\begin{equation}
\langle\dot\phi_i^2\rangle
\simeq\langle\phi_i V_{i,\phi}\rangle
=2n_i\langle V_i\rangle,
\qquad V_i\propto\phi_i^{2n_i}.
\label{eq:scalar_virial}
\end{equation}
Consequently, the averaged pressure-to-density ratio and dilution law are \cite{Turner:1983he}
\begin{equation}
\langle w_i\rangle\equiv
\frac{\langle P_{\phi_i}\rangle}{\langle\rho_{\phi_i}\rangle}
=\frac{n_i-1}{n_i+1},
\qquad
\rho_{\phi_i}\propto a^{-6n_i/(n_i+1)}.
\label{eq:averaged_dilution}
\end{equation}
For the periodic potentials this requires
\begin{equation}
\frac{|\phi_{A,i}|}{f_i}\ll1,
\qquad T_{{\rm osc},i}H\ll1,
\label{eq:averaging_conditions}
\end{equation}
where \(\phi_{A,i}\) is the oscillation amplitude and
\(T_{{\rm osc},i}\) is the period. The first condition ensures that the leading power near the
minimum describes the potential throughout the oscillation.
The second requires the background to change little during
one cycle. A field can cross the minimum before its motion
satisfies either condition.

To determine how the sextic frequency depends on amplitude,
we calculate the oscillation period, neglecting Hubble damping over a single cycle. Energy conservation then results in
\begin{align}
T_{{\rm osc},i}
&=4\int_0^{|\phi_{A,i}|}
\frac{\dd\phi}{\sqrt{2[V_i(\phi_{A,i})-V_i(\phi)]}}
\nonumber\\
&=\frac{2\sqrt{2}|\phi_{A,i}|}
{\sqrt{V_i(\phi_{A,i})}}
\int_0^1\frac{\dd y}{\sqrt{1-y^{2n_i}}},
\label{eq:monomial_period}
\end{align}
where \(y\) is the displacement divided by the amplitude, and
the second equality uses the monomial potential. Thus the
frequency scales as \(|\phi_{A,i}|^{n_i-1}\); only the quadratic case has an amplitude-independent frequency. For a sextic minimum, the shrinking amplitude lengthens the period even while the scalar energy density decreases.

In the rapid-oscillation regime, a scalar condensate near a
quadratic minimum has \(\langle w_i\rangle=0\) and
\(\rho_{\phi_i}\propto a^{-3}\), as in the standard
axion misalignment mechanism
\cite{Preskill:1982cy,Abbott:1982af,Dine:1982ah}.
For small-amplitude oscillations within the sextic minima,
the averaged equation of state and dilution law are
\begin{equation}
\langle w_i\rangle=\frac12,
\qquad\rho_{\phi_i}\propto a^{-9/2},
\qquad\frac{\rho_{\phi_i}}{\rho_r}\propto a^{-1/2},
\qquad\frac{\rho_{\phi_i}}{\rho_m}\propto a^{-3/2}.
\label{eq:n3_scaling}
\end{equation}
This distinction matters most for the earliest rung. A stable
matter-like component with a percent-level BBN fraction would
grow relative to radiation; the sextic regime instead reduces
that ratio. The suppression comes from the pressure of the
coherent field, not from decay into a thermal species.

The validity of the oscillation average depends on how the
oscillation frequency evolves relative to the Hubble rate.
At a turning point, the sextic potential satisfies
\(\rho_{\phi_i}=V_i(\phi_{A,i})\propto\phi_{A,i}^6\).
Combined with the averaged dilution law
\(\rho_{\phi_i}\propto a^{-9/2}\), this implies an
amplitude scaling of \(\phi_{A,i}\propto a^{-3/4}\). Its frequency
\(\omega_{{\rm osc},i}=2\pi/T_{{\rm osc},i}\) scales as
\(\phi_{A,i}^2\), hence as \(a^{-3/2}\). The ratio
\(\omega_{{\rm osc},i}/H\) grows during radiation domination
but stays approximately constant during matter domination
and can decrease during late acceleration. Sextic evolution during matter domination can depart from
the rapid-oscillation regime \cite{Ema:2015dza}.
We therefore determine the late residual densities from the
full scalar trajectories, using the oscillation average to
describe the motion when it is rapid and small in amplitude.

Accelerated expansion depends on the total density and
pressure of the scalar fields, matter, and radiation.
The acceleration equation is
\begin{equation}
\frac{\ddot a}{a}
=-\frac{\rho_{\rm tot}+3P_{\rm tot}}{6\Mpl^2}
=-\frac{\rho_m+2\rho_r+
2\sum_i(\dot\phi_i^2-V_i)}{6\Mpl^2}.
\label{eq:acceleration}
\end{equation}
For matter, radiation, and a nearly frozen scalar contribution \(\rho_{\rm fr}\), acceleration requires
\(2\rho_{\rm fr}>\rho_m+2\rho_r\), neglecting other scalar
contributions. With radiation alone as the competing fluid,
the frozen fraction must exceed \(1/2\); with matter alone it
must exceed \(1/3\). Rolling adds positive kinetic terms, so
these vacuum-like thresholds do not replace the exact criterion. The benchmark remains decelerating through the BBN and EDE features. Their role is to change the expansion rate over those intervals; the late-time field supplies the subsequent accelerated epoch.

% ============================================================
\section{Homogeneous evolution of the axion ladder}
\label{sec:numerical_benchmark}
% ============================================================

All four fields are evolved from \(z_{\rm ini}=10^{11}\) to the present, with zero initial velocities and the potential parameters in table~\ref{tab:benchmark}. We choose a common initial hilltop offset of \(\delta_i=0.30\) for the three transient fields and an initial angle of \(\theta_{\rm DE}=94^\circ\) for the late-time axion. The potential scale of the late-time field is adjusted to satisfy \(H(z=0)=H_0\), giving
\begin{equation}
\Lambda_{\rm DE}\simeq2.2594\times10^{-3}\,\eV.
\label{eq:lambda_de_tuned}
\end{equation}
The unrounded value is retained in the integration, as specified in appendix~\ref{app:numerics}. Closure includes the residual transient densities:
\begin{equation}
\Omega_{{\rm DE},0}
=1-\Omega_{m0}-\Omega_{r0}
-\Omega_{{\rm BBN},0}
-\Omega_{{\rm EDE1},0}
-\Omega_{{\rm EDE2},0}.
\label{eq:present_closure}
\end{equation}
We fix \(H_0=67.4\kmsMpc\) when normalizing the benchmark.

All four fields contribute to \(H\) at every integration step. The earlier-rolling EDE field contributes to the Hubble rate and thereby modifies the friction acting on the field that rolls later. Their residual densities also enter the present-day normalization.

\begin{table}[t]
\centering
\small
\caption{Effective potential parameters and calculated transient peaks. The upper block lists the input parameters, while the lower block contains the derived curvature scales and numerical results. For the three transients, \(m_{\rm eff}\) denotes the hilltop curvature scale; the late-time entries refer to the vacuum mass \(m_{\rm DE}\) and the present energy fraction. The potential scales are rounded for presentation, with the precision used for the late-time normalization specified in appendix~\ref{app:numerics}.}
\label{tab:benchmark}
\begin{tabular}{lcccc}
\toprule
Field & \(n_i\) & \(\theta_i\) & \(f_i/\Mpl\) &
\(\Lambda_i\,[\eV]\) \\
\midrule
BBN & 3 & \(\pi-0.30\) & 0.0400 &
\(5.65\times10^4\) \\
EDE1 & 3 & \(\pi-0.30\) & 0.1280 &
\(2.79\times10^{-1}\) \\
EDE2 & 3 & \(\pi-0.30\) & 0.1280 &
\(8.22\times10^{-1}\) \\
Late time & 1 & \(94^\circ\) & 1.4374 &
\(2.2594\times10^{-3}\) \\
\bottomrule
\end{tabular}

\medskip

\begin{tabular}{lccc}
\toprule
Field & \(m_{\rm eff}\) or \(m_{\rm DE}\,[\eV]\) &
\(z_{{\rm peak},i}\) &
\(\Omega_i^{\rm peak}\) or \(\Omega_{{\rm DE},0}\) \\
\midrule
BBN & \(1.14\times10^{-16}\) &
\(1.04\times10^9\) & \(9.90\times10^{-3}\) \\
EDE1 & \(8.65\times10^{-28}\) &
\(2.43\times10^3\) & \(7.81\times10^{-2}\) \\
EDE2 & \(7.51\times10^{-27}\) &
\(7.93\times10^3\) & \(8.64\times10^{-2}\) \\
Late time & \(1.46\times10^{-33}\) &
--- & \(6.85\times10^{-1}\) \\
\bottomrule
\end{tabular}
\end{table}

\begin{figure}[t]
\centering
\includegraphics[width=0.98\textwidth]{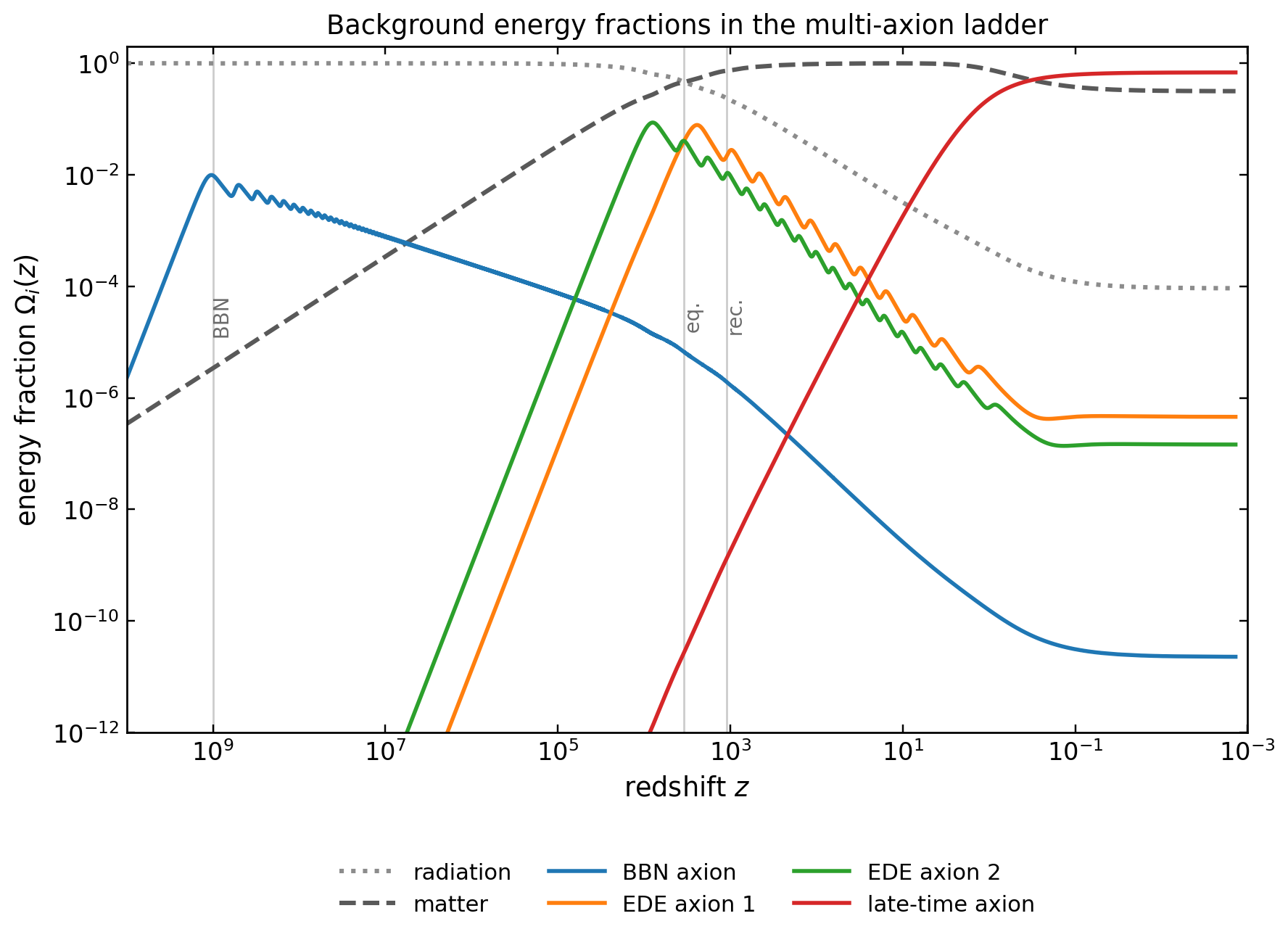}
\caption{Energy fractions in the homogeneous four-field benchmark. The solid curves show BBN, EDE1, EDE2, and late-time axion fractions; the dotted and dashed gray curves show radiation and matter. Every scalar curve is calculated from the resolved field equations, including the oscillations after the transient peaks. The vertical lines mark \(z=10^9\), matter--radiation equality at \(z_{\rm eq}\simeq3422\), and a reference recombination-era redshift \(z=1100\). The horizontal axis runs toward later times from left to right.}
\label{fig:omega_ladder}
\end{figure}

Figure~\ref{fig:omega_ladder} shows the fractional-density
histories. The heavier EDE field rolls first; the lighter EDE
field peaks after matter--radiation equality. Their combined
fraction,
\begin{equation}
\Omega_{\rm EDE,tot}(z)
\equiv\Omega_{\rm EDE1}(z)+\Omega_{\rm EDE2}(z),
\label{eq:combined_ede_fraction}
\end{equation}
reaches \(0.0971\) near \(z=2.54\times10^3\). The individual
maxima occur at different redshifts, so their sum does not equal the maximum combined contribution.

The peak locations also differ from the nominal curvature
crossings. Defining
\begin{equation}
\kappa_i\equiv
\frac{m_{{\rm eff},i}}{H(z_{{\rm peak},i})},
\label{eq:kappa}
\end{equation}
we find \(\kappa_i\simeq7.54,\ 6.52,\ 6.91\) for BBN,
EDE1, and EDE2, respectively. The background density falls
substantially between the crossing and the peak. These factors depend on the potential, initial displacement, and shared expansion history.

For comparison, the nominal crossings in the same background
occur at approximately
\begin{equation}
z_{\times,\mathrm{BBN}}=2.87\times10^9,\qquad
z_{\times,\mathrm{EDE1}}=7.02\times10^3,\qquad
z_{\times,\mathrm{EDE2}}=2.25\times10^4.
\label{eq:crossing_numbers}
\end{equation}
All three crossings precede the corresponding density peaks.
The radiation-only estimate is adequate for locating the BBN
crossing, but matter already affects the expansion near the
EDE crossings.

Table~\ref{tab:relic_safety} provides the fractions at representative epochs. The adopted fluid densities place equality at
\begin{equation}
z_{\rm eq}=\frac{\Omega_{m0}}{\Omega_{r0}}-1
\simeq3422.37.
\label{eq:equality}
\end{equation}
The BBN field has become negligible by the reference recombination epoch, whereas the two EDE fields still contribute a combined fraction of \(0.0323\). Their evolution through recombination must therefore enter any later CMB calculation. By today, the combined fraction of the three transients is only \(6.02\times10^{-7}\).

Each scalar obeys eq.~\eqref{eq:scalar_conservation}, so its
energy density decreases or remains constant during expansion. Its fractional density can nevertheless develop secondary maxima as the oscillating equation of state crosses
\(w_{\rm tot}\), according to eq.~\eqref{eq:fraction_peak_condition}. For the EDE fields,
the first few oscillations occur during the epoch relevant
to acoustic propagation.

\begin{table}[t]
\centering
\small
\caption{Calculated energy fractions at representative epochs. The equality column is evaluated at \(z_{\rm eq}=3422.37\), where the benchmark matter and radiation densities are equal, as given by eq.~\eqref{eq:equality}. The recombination-era column shows the energy fractions at the reference redshift \(z=1100\).}
\label{tab:relic_safety}
\begin{tabular}{lcccc}
\toprule
Field & \(\Omega_i(10^9)\) & \(\Omega_i(z_{\rm eq})\) &
\(\Omega_i(1100)\) & \(\Omega_{i0}\) \\
\midrule
BBN & \(9.74\times10^{-3}\) & \(6.63\times10^{-6}\) &
\(1.93\times10^{-6}\) & \(2.27\times10^{-11}\) \\
EDE1 & \(1.33\times10^{-23}\) & \(4.08\times10^{-2}\) &
\(2.17\times10^{-2}\) & \(4.56\times10^{-7}\) \\
EDE2 & \(9.99\times10^{-22}\) & \(4.04\times10^{-2}\) &
\(1.06\times10^{-2}\) & \(1.45\times10^{-7}\) \\
Late time & \(8.16\times10^{-33}\) & \(2.76\times10^{-11}\) &
\(1.32\times10^{-9}\) & \(6.85\times10^{-1}\) \\
\bottomrule
\end{tabular}
\end{table}

\begin{figure}[t]
\centering
\includegraphics[width=0.94\textwidth]{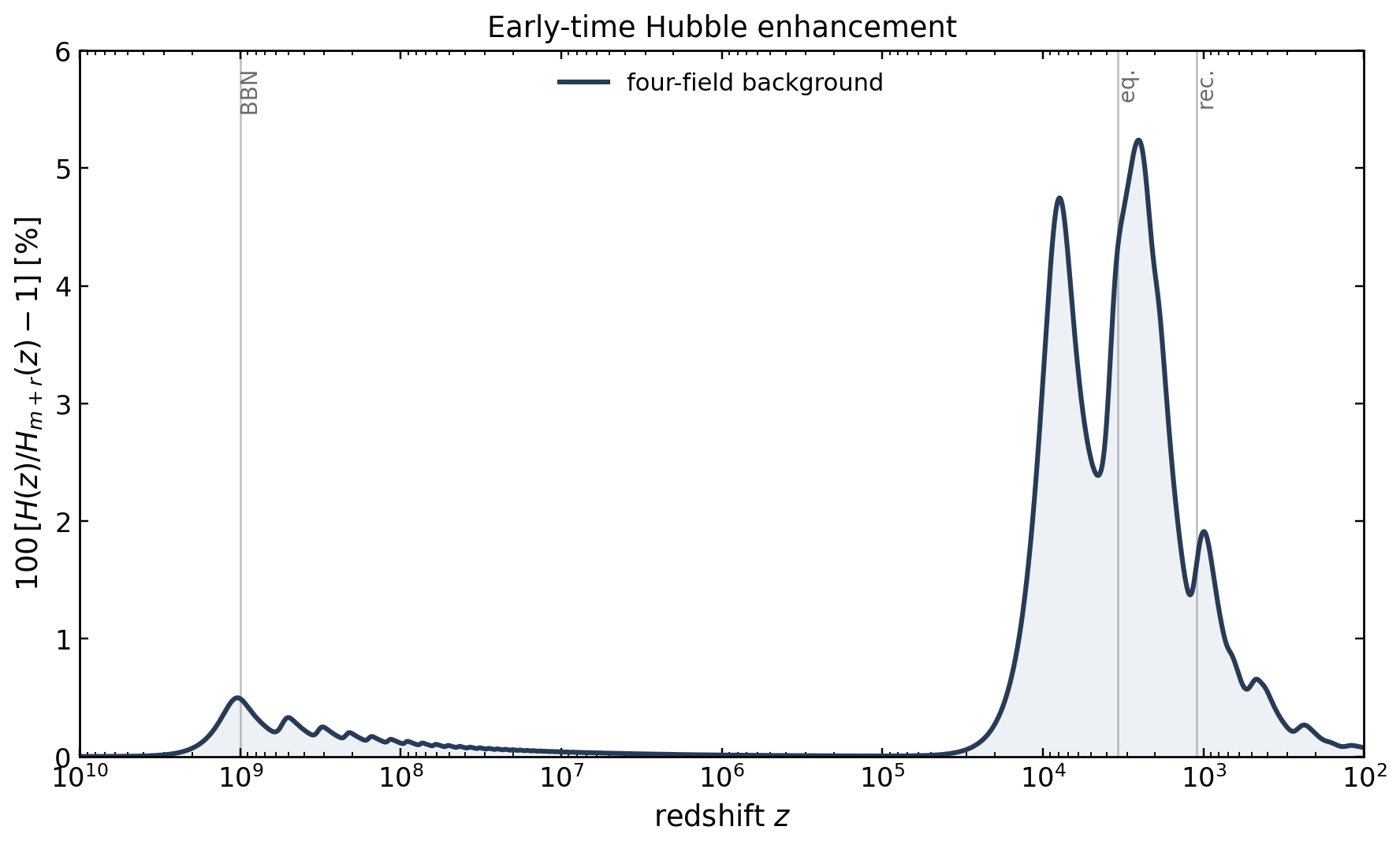}
\caption{Early-time expansion enhancement,
\(100[H(z)/H_{m+r}(z)-1]\), calculated from the same four-field solution as figure~\ref{fig:omega_ladder}. The BBN feature reaches approximately \(0.499\%\); the overlapping EDE contributions reach approximately \(5.24\%\). The structure after each roll follows from the resolved scalar oscillations. The reference rate follows eq.~\eqref{eq:h_reference}, and the vertical markers match those in figure~\ref{fig:omega_ladder}.}
\label{fig:hboost}
\end{figure}

For comparison, the matter-plus-radiation reference is defined by
\begin{equation}
H_{m+r}^2(z)=H_0^2\left[
\Omega_{m0}(1+z)^3+\Omega_{r0}(1+z)^4\right],
\qquad \Omega_{\rm scal}(z)\equiv\sum_i\Omega_i(z).
\label{eq:h_reference}
\end{equation}
From the Friedmann equation then we get the exact algebraic relation
\begin{align}
\frac{H(z)}{H_{m+r}(z)}-1
&=\left[1-\Omega_{\rm scal}(z)\right]^{-1/2}-1
\nonumber\\
&=\frac12\Omega_{\rm scal}(z)
+\frac38\Omega_{\rm scal}^2(z)
+\mathcal O(\Omega_{\rm scal}^3).
\label{eq:hboost_exact}
\end{align}
The expansion is useful for a small total scalar fraction.
The sum includes all four fields, although the late-time field contributes negligibly over the range in figure~\ref{fig:hboost}. The reference rate uses the same matter and radiation densities as the four-field solution.

At the BBN maximum, the expansion enhancement is
\(H/H_{m+r}-1\simeq0.499\%\), compared with \(0.495\%\)
in the linear approximation. The combined EDE contribution
reaches a maximum enhancement of approximately \(5.24\%\),
with values of \(4.33\%\) at exact equality and \(1.66\%\)
at \(z=1100\). The different profiles reflect the rolling
trajectories and the overlap of the two EDE fields.

Higher-order corrections become more important for the larger
EDE contribution, so all plotted values are calculated using
the exact expression.

The deceleration parameter \(q\equiv-\ddot a/(aH^2)
=-1-\dot H/H^2\) remains positive through the BBN and EDE
features. The late-time field instead drives \(q\) through
zero near \(z=0.649\), with \(q_0\simeq-0.450\).

% ============================================================
\section{Physical interpretation of the three epochs}
\label{sec:epochs}
% ============================================================

The three epochs connect the scalar evolution to different
cosmological processes. The BBN transient changes the time
available for weak and nuclear reactions. The EDE fields affect acoustic propagation, photon diffusion, and gravitational-potential evolution before recombination, while the late-time field changes cosmological distances and the growth of structure. These effects depend on the timing and duration of each contribution as well as its peak fraction.

\subsection{The BBN transient}

The first field produces the near-percent fractional-density peak and approximately half-percent expansion enhancement shown in figure~\ref{fig:hboost}. It does so without transferring energy to the visible plasma. While Hubble friction holds the field nearly fixed, radiation dilution increases its relative importance. Its subsequent motion removes most of that contribution, leaving the small residuals in table~\ref{tab:relic_safety}. This history
differs from adding a constant radiation fraction: the scalar's pressure changes as it rolls, and its fractional density varies across the relevant interval.

To locate the transient relative to weak freeze-out and the
deuterium bottleneck, the BBN calculation will evolve the
photon temperature together with the scalar background.

Weak interactions first control the neutron-to-proton ratio.
Neutron decay then changes that ratio before nuclei form
efficiently. Once deuterium survives photodissociation, nuclear reactions both produce and destroy it, and assemble most surviving neutrons into helium-4. A change in expansion can affect the time available at each stage. Helium therefore responds not only to weak freeze-out but also to the subsequent survival of neutrons, while deuterium probes the later competition between nuclear burning and expansion
\cite{Pitrou:2018cgg,Fields:2019pfx,Cyburt:2015mya}.

The effect on light-element abundances depends on when the
expansion is enhanced relative to weak freeze-out and nuclear
burning, and on how long the enhancement persists
\cite{Cook:2025gra,Poulin:2026ltf}. A vacuum-like contribution that subsequently converts into radiation or kination has been considered in earlier BBN studies
\cite{McKeen:2024voa}. In our model, the axion remains
nearly frozen until Hubble damping weakens sufficiently
for it to roll. Its subsequent evolution is determined
by the sextic minimum: rapid, small-amplitude oscillations
give \(\langle w_\phi\rangle=1/2\) and
\(\rho_\phi\propto a^{-9/2}\), allowing the field to dilute
faster than radiation without transferring energy to the
thermal plasma. The same scalar dynamics therefore
determine both the expansion enhancement during BBN and
the residual density at later epochs.

\subsection{Two-field early dark energy}

The heavier EDE field reaches its maximum before
matter--radiation equality, while the lighter field peaks
afterward. Their overlapping contributions raise the Hubble
rate over a broader redshift interval than either field alone. The combined profile retains two main peaks, together with smaller variations produced by the subsequent scalar
oscillations. The peak separation and the dilution of each
field therefore determine both the duration and the shape
of the expansion enhancement before recombination.

The comoving distance traveled by acoustic waves defines
the sound horizon. We distinguish its values at baryon drag
and photon decoupling
\cite{Hu:1995en,Eisenstein:1997ik}:
\begin{equation}
r_s(z)=\int_z^\infty
\frac{c_s(z')}{H(z')}\,\dd z',
\qquad
r_d\equiv r_s(z_d),
\qquad
r_*\equiv r_s(z_*),
\label{eq:sound_horizons}
\end{equation}
where \(c_s\) is the photon--baryon sound speed,
\(z_d\) is the baryon-drag redshift, and \(z_*\) is the
photon-decoupling redshift. In the tight-coupling limit,
photon pressure and the inertia of both species determine
the sound speed \cite{Eisenstein:1997ik}:
\begin{equation}
c_s^2=\frac{1}{3(1+R_b)},
\qquad
R_b\equiv\frac{3\rho_b}{4\rho_\gamma},
\label{eq:sound_speed}
\end{equation}
where \(R_b\) measures the baryon loading, and \(\rho_b\)
and \(\rho_\gamma\) are the baryon and photon energy densities.

The drag horizon calibrates BAO distances, whereas the CMB
acoustic angle relates the sound horizon at photon decoupling
to its apparent angular size. In the assumed spatially flat
background, these relations take the form
\cite{Planck:2018vyg,Hogg:1999ad}
\begin{equation}
\theta_*=\frac{r_*}{D_M(z_*)},
\qquad
D_M(z)=\int_0^z\frac{\dd z'}{H(z')}
      =(1+z)D_A(z),
\label{eq:acoustic_angle}
\end{equation}
where \(D_M\) is the transverse comoving distance and
\(D_A\) is the physical angular-diameter distance.

To isolate the expansion response, consider a change
\(\Delta H(z)\) with \(|\Delta H/H|\ll1\), holding
\(c_s(z)\) and \(z_*\) fixed. Let \(\Delta r_*\) denote
the corresponding change in the sound horizon.
Expanding eq.~\eqref{eq:sound_horizons} to first order we get
\begin{equation}
\begin{aligned}
\frac{\Delta r_*}{r_*}
&\simeq-\int_{z_*}^{\infty}W_*(z)
\frac{\Delta H(z)}{H(z)}\,\dd z,
\\
W_*(z)&\equiv\frac{c_s(z)}{H(z)r_*},
\qquad
\int_{z_*}^{\infty}W_*(z)\,\dd z=1.
\end{aligned}
\label{eq:sound_horizon_response}
\end{equation}
Here \(H\), \(c_s\), and \(r_*\) refer to the unperturbed
background, and \(W_*\) is the resulting normalized acoustic
weight. The same expression describes the drag-horizon response with \(z_*\) and \(r_*\) replaced by \(z_d\) and \(r_d\), respectively. At fixed sound speed and endpoints, a positive expansion change reduces the horizon according to this weighted integral. The effect therefore depends on when the EDE contribution occurs and how long it lasts, as well as on its peak fraction. The full response also includes changes in the thermal history and the decoupling and drag redshifts \cite{Poulin:2018cxd,Smith:2019ihp}.

Photon diffusion provides a separate constraint. Its length
depends on the scattering mean free path and the time available for a random walk, not simply on the distance traveled by a sound wave \cite{Hu:1995en}. Even at fixed ionization history, diffusion and sound propagation weight expansion differently. The acoustic angle can therefore remain unchanged while photon diffusion suppresses CMB anisotropies at a different angular scale. The fields also change the gravitational potentials that drive acoustic oscillations, so their pressure and density perturbations affect peak heights and the temperature signal from time-varying potentials \cite{Ma:1995ey,Smith:2019ihp}.

\subsection{Late-time thawing dark energy}

The late-time field has a vacuum mass of order \(H_0\).
Hubble friction keeps it nearly frozen through most of cosmic
history, so its potential energy changes slowly while matter
dilutes. Its fractional contribution therefore grows toward
the present epoch. As the expansion rate falls, the field
begins to roll and develops kinetic energy, raising its
equation of state above \(-1\), as expected for thawing
quintessence
\cite{Frieman:1995pm,Caldwell:2005tm,Scherrer:2007pu,Dutta:2008qn,Kaloper:2005aj}.
The field remains potential dominated and has not entered
the rapid-oscillation regime reached by the early transients.
Its present energy fraction and equation of state are
\begin{equation}
\Omega_{{\rm DE},0}\simeq0.6849,
\qquad
w_{\rm DE}(a=1)\simeq-0.9249.
\label{eq:late_present}
\end{equation}
Thus the field supplies the dominant present energy component, with a density that decreases slowly as it rolls.

For comparison with common descriptions of late-time expansion, we fit the Chevallier--Polarski--Linder (CPL) form
\cite{Chevallier:2000qy,Linder:2002et},
\begin{equation}
w_{\rm DE}(a)\simeq
w_0^{\rm fit}+w_a^{\rm fit}(1-a).
\label{eq:cpl}
\end{equation}
Fitting \(w_{\rm DE}(a)\) to the CPL form by least squares
over \(0\leq z\leq2\), with equal weight per unit
\(\ln a\), we obtain
\begin{equation}
w_0^{\rm fit}\simeq-0.9295,
\qquad
w_a^{\rm fit}\simeq-0.1051.
\label{eq:w0wa_result}
\end{equation}
The negative fitted \(w_a\) reflects an equation of state
closer to \(-1\) at higher redshift, followed by an increase
as the field thaws. The root-mean-square residual is
\(2.26\times10^{-3}\). The fitted intercept differs slightly
from \(w_{\rm DE}(a=1)\) because the linear CPL form
approximates the curved scalar trajectory over the full
fitted interval.

The canonical kinetic term and positive scalar energy
density imply
\begin{equation}
1+w_{\rm DE}
=\frac{\dot\phi_{\rm DE}^{\,2}}
{\rho_{\phi_{\rm DE}}}
\geq0.
\label{eq:nonphantom}
\end{equation}
This condition holds throughout the scalar evolution,
whereas the fitted CPL coefficients describe only the
specified redshift interval. Comparisons with DESI-era
scalar-field analyses
\cite{Berghaus:2024kra,Wolf:2024eph,Bhattacharya:2024kxp,Shlivko:2024llw}
therefore require the full trajectory, which determines
the dark-energy density and its contribution to the
expansion history in a joint cosmological fit.

% ============================================================
\section{Potential and initial-condition requirements}
\label{sec:consistency}
% ============================================================

The four-field calculation assumes negligible energy exchange
with the visible plasma and transient potentials with sextic
minima. We examine the direct couplings and inflationary initial conditions associated with this evolution, including fluctuations around the adopted misalignments. We then consider how quadratic and quartic corrections could change the subsequent dilution of the BBN and EDE fields.

\subsection{Direct couplings}

The large decay constants suppress conventional axion couplings when their dimensionless coefficients are of order unity. For example, the photon interaction can be written as
\begin{equation}
\mathcal L_{\phi_i\gamma\gamma}
=-\frac14g_{\phi_i\gamma\gamma}\phi_i
F_{\mu\nu}\widetilde F^{\mu\nu},
\qquad
g_{\phi_i\gamma\gamma}
=C_{i\gamma}\frac{\alpha}{2\pi f_i},
\label{eq:photon_coupling}
\end{equation}
where \(F_{\mu\nu}\) is the electromagnetic field strength,
\(\widetilde F^{\mu\nu}\) its dual, \(\alpha\) the fine-structure constant, and \(C_{i\gamma}\) a model-dependent anomaly coefficient. For \(C_{i\gamma}=1\), the BBN, either EDE, and late-time couplings are approximately
\begin{equation}
\left(1.19\times10^{-20},\quad
3.73\times10^{-21},\quad
3.32\times10^{-22}\right)\,\GeV^{-1},
\label{eq:coupling_values}
\end{equation}
respectively. These values lie far below conventional laboratory and stellar-cooling sensitivities
\cite{Raffelt:2006cw,Caputo:2024oqc,Jaeckel:2010ni,
Irastorza:2018dyq}.

The anomaly coefficient is an independent microscopic input.
The values in eq.~\eqref{eq:coupling_values} assume
\(C_{i\gamma}=1\); enhanced coefficients or additional
interactions require a separate assessment.

A derivative interaction with a fermion \(\psi\) has the form
\begin{equation}
\mathcal L_{\phi_i\psi}
=\frac{C_{i\psi}}{2f_i}\partial_\mu\phi_i\,
\bar\psi\gamma^\mu\gamma^5\psi,
\label{eq:fermion_coupling}
\end{equation}
where \(C_{i\psi}\) is its dimensionless coefficient; the
corresponding on-shell pseudoscalar coupling scales as
\(C_{i\psi}m_\psi/f_i\), with \(m_\psi\) the fermion mass.
We take the axions to remain nonthermal, with negligible energy exchange with the visible plasma. We also neglect non-derivative scalar couplings to ordinary matter. For nonrelativistic fermions, single-pseudoscalar exchange
through these derivative couplings produces a spin-dependent
interaction, in contrast to the spin-independent force generated by a scalar coupling to the trace of the matter stress tensor.

The transient hilltop scales cannot be inserted into the usual massive-boson black-hole superradiance limits, because their pure sextic minima have no quadratic vacuum mass
\cite{Arvanitaki:2010sy,Brito:2015oca}. Such a test would require the curvature about the relevant background and any corrections to the potential. The late-time field has a vacuum mass, but its scale is far below the standard stellar-mass black-hole range.

\subsection{Inflationary initial conditions and isocurvature}
\label{subsec:isocurvature}

For the benchmark, we adopt
\(\delta_{\rm BBN}=\delta_{\rm EDE1}=\delta_{\rm EDE2}=0.30\)
and \(\theta_{\rm DE}=94^\circ\). The initial angles are independent inputs; the common-offset scan in appendix~\ref{app:hilltop} varies the three transient angles together to measure their joint effect on the density peaks. These angles set the initial potential energies, while the overdamping estimates in appendix~\ref{app:inflation} motivate small initial velocities. We assume that inflation-dependent potential terms and thermal or nonthermal interactions preserve the adopted displacements
through reheating.

A light axion during inflation acquires fluctuations in its
initial misalignment. Let \(H_{\rm inf}\) denote the inflationary Hubble rate when the pivot mode \(k_*\) exits the horizon. For a canonical field with \(|V_{i,\phi\phi}|\ll H_{\rm inf}^2\),
the dimensionless power spectrum \(\mathcal P_{\delta\theta_i}\)
of its angular fluctuation \(\delta\theta_i\) satisfies
\begin{equation}
\mathcal P_{\delta\theta_i}^{1/2}(k_*)
\simeq\frac{H_{\rm inf}}{2\pi f_i}.
\label{eq:isocurvature_fluct}
\end{equation}
We take the axions to be weakly coupled to the inflaton, with
initially uncorrelated inflaton and axion fluctuations
\cite{Marsh:2015xka,Planck:2018jri}.

Using the inflationary normalization in
appendix~\ref{app:inflation}, we obtain angular fluctuation
amplitudes of approximately
\begin{equation}
\left(
4.1\times10^{-5},\quad
1.3\times10^{-5},\quad
1.1\times10^{-6}
\right)
\label{eq:angular_fluctuations}
\end{equation}
for the BBN field, each of the two EDE fields, and the
late-time field, respectively. We evaluate the initial perturbations on a common spatially flat hypersurface. Neglecting the fields' initial kinetic energy, we obtain the linear density perturbations
\begin{equation}
\left.\frac{\delta\rho_{\phi_i}}{\bar\rho_{\phi_i}}
\right|_{\rm ini}
\simeq n_i\frac{\sin\theta_i}{1-\cos\theta_i}\,
\delta\theta_i,
\label{eq:initial_density_response}
\end{equation}
where the bar denotes the homogeneous density.

For the common transient angle, the coefficient multiplying
\(\delta\theta_i\) is \(0.453\). The corresponding initial
fractional-density amplitudes are approximately
\(1.84\times10^{-5}\) for BBN and \(5.74\times10^{-6}\)
for either EDE field; the late-time value is approximately
\(1.05\times10^{-6}\). These amplitudes quantify the density perturbations of the initially frozen fields in the spatially flat hypersurface. The same angular fluctuations also shift the onset of rolling. Near the hilltop, the rolling delay and the frozen potential energy have different sensitivities to the initial misalignment, so a small initial density contrast alone does not determine
the subsequent perturbation amplitude.

The angular fluctuations also change the later rolling
trajectories. The common-offset scan in appendix~\ref{app:hilltop}
measures the joint response to the three transient angles.
At fixed potential parameters,
\begin{equation}
\frac{\dd\ln\Omega_i^{\rm peak}}{\dd\ln\delta}
=\left.
\sum_{j\in\{\mathrm{BBN},\mathrm{EDE1},\mathrm{EDE2}\}}
\frac{\partial\ln\Omega_i^{\rm peak}}{\partial\ln\delta_j}
\right|_{\delta_j=\delta}.
\label{eq:common_offset_response}
\end{equation}
Each derivative includes the gravitational coupling between
the fields through the common Friedmann background. Varying
one initial angle at a time, with the others held fixed,
isolates the response to each independent misalignment
fluctuation. These derivatives describe how the homogeneous
peak fractions and their epochs respond to the initial
conditions. Predicting the CMB isocurvature signal also
requires the coupled evolution of the field, metric,
radiation, and matter perturbations, initialized on a
common spatially flat hypersurface.

\subsection{Sextic minima and corrections to the potential}

Approximate shift symmetry can protect the small overall
scale of an axion potential. The sextic minima of the BBN
and EDE fields require a further relation among the first
three harmonics:
\begin{equation}
[1-\cos\Theta_i]^3
=\frac52-\frac{15}{4}\cos\Theta_i
+\frac32\cos(2\Theta_i)-\frac14\cos(3\Theta_i).
\label{eq:sextic_harmonics}
\end{equation}
Expanding about \(\Theta_i=0\), these coefficients cancel
the quadratic and quartic terms, leaving
\(\Theta_i^6/8\) as the leading contribution. The overall
potential scale therefore controls the strength of the
symmetry breaking, while the relative harmonic amplitudes
and phases determine the shape of the minimum.

In a microscopic description, the sextic minimum follows from
the relative amplitudes and phases of the contributing harmonics \cite{Czerny:2014wza,McDonough:2022pku,Cicoli:2023qri}.
Additional explicit symmetry breaking can modify these relations and generate quadratic or quartic terms near the minimum \cite{Kamionkowski:1992mf}. These contributions decrease more slowly with field amplitude than the sextic term, so even small corrections can dominate the later evolution. If this occurs well after the density peak, the early transient can remain nearly unchanged while the subsequent dilution and surviving density are altered.

To state the required suppression near the minimum, consider
\begin{equation}
V_i=\Lambda_i^4\left[
\frac{\Theta_i^6}{8}+c_{4,i}\Theta_i^4+
\mathcal O(\Theta_i^8)\right]
+\frac12\mu_i^2f_i^2\Theta_i^2,
\label{eq:lower_power_corrections}
\end{equation}
where \(c_{4,i}\) is a dimensionless quartic coefficient and
\(\mu_i\) is a possible non-negative quadratic mass. The
benchmark has \(c_{4,i}=\mu_i=0\). At angular amplitude
\(\Theta_{A,i}\equiv|\phi_{A,i}|/f_i\), sextic
potential-energy dominance requires
\begin{equation}
\frac{\mu_i f_i}{\Lambda_i^2}
\ll\frac{\Theta_{A,i}^2}{2},
\qquad
|c_{4,i}|\ll\frac{\Theta_{A,i}^2}{8}.
\label{eq:sextic_dominance}
\end{equation}
The corrected potential must also retain a stable minimum.

During rapid oscillations, a dominant quartic or quadratic term gives an averaged density scaling of \(a^{-4}\) or \(a^{-3}\), respectively. A quadratic tail therefore grows relative to radiation and remains approximately constant relative to matter. Determining how the correction changes the residual fraction in table~\ref{tab:relic_safety} requires evolving the field with the corrected potential through the crossover and up to the present.

An effective decay constant larger than \(\Mpl\) is adopted
for the late-time axion. Such an enhanced range can be
generated through alignment among periodic directions with
smaller microscopic decay constants. The resulting shallow
potential must, however, be assessed with all relevant
instanton contributions included, since its slope and
curvature can be modified by additional harmonics.

Restrictions on axion decay constants and instanton actions
have been derived under proposed quantum-gravity conditions,
including the axion weak gravity conjecture. Their implications for EDE and axion quintessence have been investigated in refs.~\cite{Rudelius:2022gyu,Shiu:2026edl}. The microscopic consistency of an aligned construction must therefore be examined through its instanton spectrum, rather than inferred from the enhanced effective range alone.

% ============================================================
\section{Cosmological tests of the axion ladder}
\label{sec:predictions}
% ============================================================

Future work will test the model through quantitative comparisons with cosmological observations. It has two parts: a consistent BBN abundance calculation and a joint CMB, BAO, and supernova analysis of the early- and late-time fields. The central EDE question is whether distributing the contribution between two fields improves the full likelihood relative to conventional single-field EDE.

\subsection{Light-element abundances}

A future BBN calculation will evolve the scalar fields alongside the thermal plasma, determining the expansion rate and the relation between photon temperature and scale factor
self-consistently. The additional scalar density changes the
time available for weak reactions, neutron decay, and nuclear
burning. A reaction network such as PRIMAT or AlterBBN will
quantify the resulting changes in the light-element abundances \cite{Pitrou:2018cgg,Arbey:2018zfh}. Following the full expansion history will distinguish effects on neutron survival from those arising during deuterium burning, since these processes respond to changes in expansion at different temperatures
\cite{Cook:2025gra}.

To isolate the scalar contribution, we will compare the reference and ladder cosmologies using the same baryon density, neutron lifetime, nuclear rates, and neutrino treatment. We will test deuterium and helium-4 jointly, including observational uncertainties and uncertainties in the nuclear rates. We will use the same baryon density in the BBN and CMB calculations and include the predicted helium abundance when computing the free-electron density during recombination. This treatment will connect the abundance and CMB predictions within the same cosmology.

\subsection{Single-field and two-field EDE in the CMB likelihood}

The central question is whether distributing the EDE contribution between two fields improves the fit to CMB observations relative to a conventional single-field model. In our benchmark, the fields peak on opposite sides of matter--radiation equality, allowing their relative contributions to change the expansion over different intervals. This freedom affects both acoustic propagation and the gravitational potentials that drive the photon--baryon oscillations. Two histories with similar sound horizons can therefore produce different acoustic peak heights and polarization spectra \cite{Smith:2019ihp}.

The perturbations also depend on the shape of each potential.
For the third-power cosines, the curvature changes along the
trajectory as the field leaves the hilltop and oscillates
about the minimum. This time-dependent curvature enters the
field perturbation equations and, through their coupling to
the metric, affects the gravitational response. The CMB thus
tests the scalar dynamics as well as the extra background
density. Photon diffusion, CMB lensing, and matter clustering
provide further constraints because they respond differently
to the expansion and perturbation histories
\cite{Hill:2020osr,Ivanov:2020ril,DAmico:2020ods,
Smith:2020rxx,Simon:2022adh,McDonough:2023qcu}.

A quantitative comparison requires the scalar and metric
perturbations to evolve together with the ordinary cosmological components. A future analysis using the Cosmic Linear Anisotropy Solving System (CLASS) or Code for Anisotropies in the Microwave Background (CAMB) can connect the backgrounds calculated here to the temperature, polarization, and lensing spectra
\cite{lesgourgues2011cosmiclinearanisotropysolving,
Diego_Blas_2011,Lewis:1999bs}.

The additional field introduces more freedom in the timing,
amplitude, and duration of the EDE contribution. Any improvement in the best-fit likelihood must therefore be assessed alongside this greater flexibility, with the same observations and late-time assumptions used for both models \cite{Herold:2021ksg}.
The decisive test is whether the broader two-field contribution better reproduces the observed CMB spectra once these effects are taken into account.

\subsection{Late-time distances and growth}

As the late-time axion thaws, its kinetic energy becomes more
important and its equation of state departs from \(-1\).
The resulting density evolution changes the expansion rate
and the distances to a given redshift
\cite{Caldwell:2005tm,Scherrer:2007pu}. Distance predictions
therefore depend on the scalar trajectory, while the CPL fit
provides a compact description of its equation of state over
the fitted redshift range. Supernovae constrain relative
luminosity distances, whereas anisotropic BAO measurements
constrain the transverse combination \(D_M(z)/r_d\) and the
radial combination \(1/[H(z)r_d]\)
\cite{Brout:2022vxf,DESI:2025zgx}.

These observables connect the early and late contributions
in the ladder. The EDE fields alter the sound horizon before
baryon drag, while the thawing field affects the subsequent
expansion. For a fixed measured value of \(D_M/r_d\), a smaller sound horizon implies a smaller inferred comoving distance. BAO therefore constrains the early acoustic ruler and late-time expansion together. Supernovae help distinguish these effects by constraining the redshift dependence of distances through a relative calibration that is independent of \(r_d\). The combined distance constraints consequently depend on both the EDE evolution and the late-time scalar trajectory.

Structure growth provides a further connection between the
epochs. On scales well inside the Hubble radius, pressure
support keeps the canonical late-time axion comparatively
smooth. Its main influence on matter clustering comes through
the expansion rate and the changing matter fraction
\cite{Copeland:2006wr}. The EDE fields also affect the matter
transfer function through their earlier influence on expansion and gravitational-potential evolution. This imprint on the amplitude and shape of matter clustering can persist after their background densities have become negligible \cite{Ivanov:2020ril,DAmico:2020ods}. Redshift-space distortions probe the peculiar velocities associated with growth, while lensing responds to the intervening gravitational potentials and cosmological distances.

The field parameters are independent inputs, but these
observational effects link them across cosmic history.
A parameter choice that accounts for the distance measurements also specifies an acoustic calibration and a growth history. Their agreement with observations provides a joint test of the early and late components of the ladder.

\subsection{Searching for additional rungs}
\label{subsec:rung_search}

The tests of primordial light-element abundances, CMB
anisotropies, and late-time distances and growth discussed
above address the fields selected for our benchmark. The ladder hypothesis also raises a broader question: at what other epochs could an axion contribute appreciably to the expansion? Rather than choosing those epochs from existing observational tensions, we can ask how much additional expansion observations permit throughout cosmic history. Such a search would distinguish tightly constrained intervals from those that still allow larger transient contributions.

BBN and recombination provide precise tests of early expansion, while BAO, supernovae, and structure growth constrain the recent Universe. An additional axion contribution between recombination and the onset of late-time acceleration could modify the expansion
during matter domination and leave an imprint on the growth
of structure. The matter power spectrum and the Lyman-\(\alpha\) forest probe the growth and scale dependence of density fluctuations \cite{Hlozek:2014lca,Goldstein:2023gnw}, while CMB lensing
responds to the intervening gravitational potentials over
a broad redshift range \cite{ACT:2023kun}. High-redshift BAO
and the forest Alcock--Paczy\'nski effect constrain distances
and expansion \cite{DESI:2025zgx,DESI:2026lnd}. Cosmic
chronometers provide complementary information through
differential galaxy ages \cite{Jimenez:2001gg}.
Future 21-cm observations may extend this sensitivity into
the cosmic dark ages and cosmic dawn, with the thermal and
ionization histories included in the interpretation
\cite{Furlanetto:2006jb}.

We describe a localized increase in the expansion rate using
a profile with an adjustable central redshift, width, and amplitude:
\begin{equation}
\frac{H^2(z)-H_{\rm ref}^2(z)}{H_{\rm ref}^2(z)}
=A_{\rm tr}\exp\left[
-\frac{\left\{\ln[(1+z)/(1+z_c)]\right\}^2}{2\sigma^2}
\right],
\qquad A_{\rm tr}\geq0,\quad \sigma>0.
\label{eq:rung_template}
\end{equation}
Here \(H_{\rm ref}(z)\) is the reference expansion history,
\(z_c\) is the center of the excess, and \(\sigma\) measures
its width in \(\ln(1+z)\). The amplitude \(A_{\rm tr}\)
is the fractional increase in \(H^2\) at the center,
corresponding to an increase in \(H\) of
\(\sqrt{1+A_{\rm tr}}-1\). With the reference component
densities held fixed, the extra energy fraction at the
center is \(A_{\rm tr}/(1+A_{\rm tr})\).
The Gaussian provides a convenient description of the
feature without selecting an axion potential in advance.

For each choice of \(z_c\) and \(\sigma\), observations
would place an upper bound on \(A_{\rm tr}\), allowing
the other cosmological parameters to adjust.
Repeating this comparison across redshift would determine
the allowed amplitude as a function of epoch for each
transient duration. General BBN expansion-history constraints
already illustrate how observations can restrict departures
from a reference history \cite{Cook:2025gra}.
Extending this question across cosmic time would show
where further rungs remain observationally possible.
Distances and structure growth depend on both the magnitude
and duration of the change in expansion. Observational bounds
on the peak amplitude therefore depend on the width of the
transient.

Equation~\eqref{eq:rung_template} describes the background
part of this search. For lensing and clustering, the
physical component producing the excess also determines
the evolution of perturbations. Axion trajectories offer
a direct way to make this connection: their potentials
specify both the allowed expansion profiles and the
scalar perturbation dynamics. The third-power cosine
fields studied here therefore provide a family of
physical transients against which the empirical bounds
can be compared, including their asymmetric rolling
and dilution histories.

For an expansion feature reproduced by an axion trajectory,
the relation between its peak epoch and curvature scale is
\begin{equation}
m_{{\rm eff},i}
=\kappa_i H(z_{{\rm peak},i}),
\label{eq:rung_mass_translation}
\end{equation}
where \(\kappa_i\) is the rolling-delay factor defined
in eq.~\eqref{eq:kappa}. Its benchmark values range
from \(6.52\) to \(7.54\), showing why the peak epoch
requires a more careful translation than simply setting
\(m_{{\rm eff},i}=H\). Accounting for this delay for
specified potentials would connect the epoch of a
feature to the axion curvature scale, while its amplitude
and duration would further constrain the decay constant
and initial angle. The expansion history can therefore
test the contributions studied here and constrain
additional axion fields that could become important
at other epochs.
% ============================================================
\section{Discussion and conclusions}
\label{sec:discussion}
\label{sec:conclusions}
% ============================================================
We have examined how four axion-like fields can contribute at
widely separated epochs within a single homogeneous cosmology. One produces a BBN-era transient, two contribute around matter--radiation equality, and the fourth drives late-time acceleration. All four fields are evolved together, so the expansion generated by their combined energy density also determines the Hubble damping experienced by each field. The calculation follows their departure from the initially frozen state, the overlap of the EDE contributions, and the subsequent evolution of the transient densities to the present. The aligned sector connects this construction to inflation through estimates of the effective field range, initial fluctuations, and post-inflationary damping.

The distinction between the early peaks and the surviving
densities is important. The peak of each transient depends
on the energy stored in the displaced field and on the time
spent near the potential maximum. Its residual density is
determined by the subsequent motion as the Universe passes
through radiation, matter, and dark-energy domination.
The sextic minima allow faster-than-radiation dilution during
rapid, small-amplitude oscillations, but the amount of
dilution follows from the resolved trajectories. This becomes
particularly relevant when the oscillation frequency is no
longer large compared with the Hubble rate, so that the
oscillation average no longer describes the evolution
reliably.

The parameter dependence clarifies how these histories are
related to the underlying axion scales. For each transient,
we can write
\begin{equation}
\Lambda_i^4=\frac{m_{{\rm eff},i}^2f_i^2}{12},
\label{eq:parameter_basis}
\end{equation}
and specify the model through
\((m_{{\rm eff},i},f_i,\theta_i)\) at fixed \(n_i=3\).
At fixed curvature scale and initial angle, the frozen
potential energy is proportional to \(f_i^2\). Changing
the initial offset affects both this energy and the time
required to leave the hilltop. Once a field contributes
appreciably to the expansion, it also changes the damping
of the other fields. The peak locations and amplitudes
therefore reflect their coupled evolution, even though
the potential parameters and initial angles are specified
independently.

The common-offset scan in appendix~\ref{app:hilltop}
illustrates this dependence. Varying the initial displacement
while holding the decay constants and potential scales fixed
changes the peak fractions more strongly than the peak
redshifts over the sampled range. The hierarchy of curvature
scales sets the characteristic epochs, while the rolling
delays shift the actual maxima. An observed expansion
feature would therefore constrain the curvature scale
together with this delay; its amplitude and duration would
supply further information about the decay constant and
initial displacement.

The surviving densities also connect the cosmological
evolution to the microscopic form of the potentials.
The sextic minimum follows from the relative amplitudes
and phases of the contributing harmonics
\cite{Czerny:2014wza,McDonough:2022pku,Cicoli:2023qri}.
Quadratic or quartic corrections decrease more slowly
with field amplitude than the sextic contribution and
can become dominant closer to the minimum. If this happens
well after the first density peak, the early transient
can remain nearly unchanged while its later dilution
is altered. The crossover conditions discussed in
section~\ref{sec:consistency} identify when such corrections
become relevant to the calculated remnants.

Future work will examine the light-element abundances through
a thermally consistent BBN reaction-network calculation and
assess the model through a joint analysis of CMB, BAO, and
supernova observations. For BBN, the
relevant question is how the time-dependent expansion change
affects neutron survival and nuclear processing. For EDE,
the two staggered fields distribute the extra energy across
matter--radiation equality and modify both acoustic propagation and gravitational-potential evolution. Evolving their scalar and metric perturbations will establish whether this freedom improves the full CMB likelihood relative to conventional single-field EDE, after accounting for the additional parameters. 

The ladder hypothesis further motivates a search beyond
the epochs selected for this benchmark. As discussed in
section~\ref{subsec:rung_search}, observations can constrain
the amplitude of a transient contribution as a function
of its epoch and duration. The interval between recombination
and late-time acceleration is of particular interest because
distance measurements and probes of structure provide
different sensitivities across this period. For a specified
potential, accounting for the rolling delay relates the
epoch of an expansion feature to the axion curvature scale,
while its amplitude and duration further constrain the
decay constant and initial angle. The expansion history
can therefore test the contributions studied here and
constrain additional axion fields that could become
important at other epochs.

\begin{acknowledgments}
AS and DS are partially supported by the U.S. National Science Foundation (NSF) under Grant No.~PHY-2310363. AS also acknowledges support under NSF Grant No.~OAC-2417682 and from a Universities Research Association Visiting Scholars Fellowship at Fermilab. We acknowledge the use of ChatGPT (OpenAI) for assistance with writing and sentence-level editing.
\end{acknowledgments}

\appendix

% ============================================================
\section{Hilltop sensitivity of the transient peaks}
\label{app:hilltop}
% ============================================================

In this appendix, we examine how the initial misalignments
affect the peak fractions and peak redshifts of the BBN and
EDE fields. By holding the potential parameters fixed, we
isolate the dependence on the initial displacements and
the associated rolling delays. The scan tests how strongly
the benchmark peak amplitudes and the ordering of the
transient epochs depend on the adopted initial angles.

The three transient offsets are varied together:
\begin{equation}
\delta_{\rm BBN}=\delta_{\rm EDE1}
=\delta_{\rm EDE2}\equiv\delta,
\qquad 0.05\leq\delta\leq0.80.
\label{eq:common_offset}
\end{equation}
All decay constants, potential scales, and the late-time
initial angle remain fixed at their benchmark values,
including \(\Lambda_{\rm DE}\). The benchmark corresponds
to \(\delta=0.30\).

The linearized equation~\eqref{eq:hilltop_linearized}
explains the dependence on the initial offset. A field
starting closer to the maximum requires a larger
amplification of its displacement before nonlinear rolling
begins. During this additional delay, its potential energy
remains nearly constant while matter and radiation dilute,
allowing its fractional contribution to grow.A smaller hilltop offset also corresponds to a larger
initial potential energy. Both effects favor a higher peak, whose amplitude and redshift are determined by the full nonlinear evolution.

\begin{figure}[t]
\centering
\includegraphics[width=0.90\textwidth]
{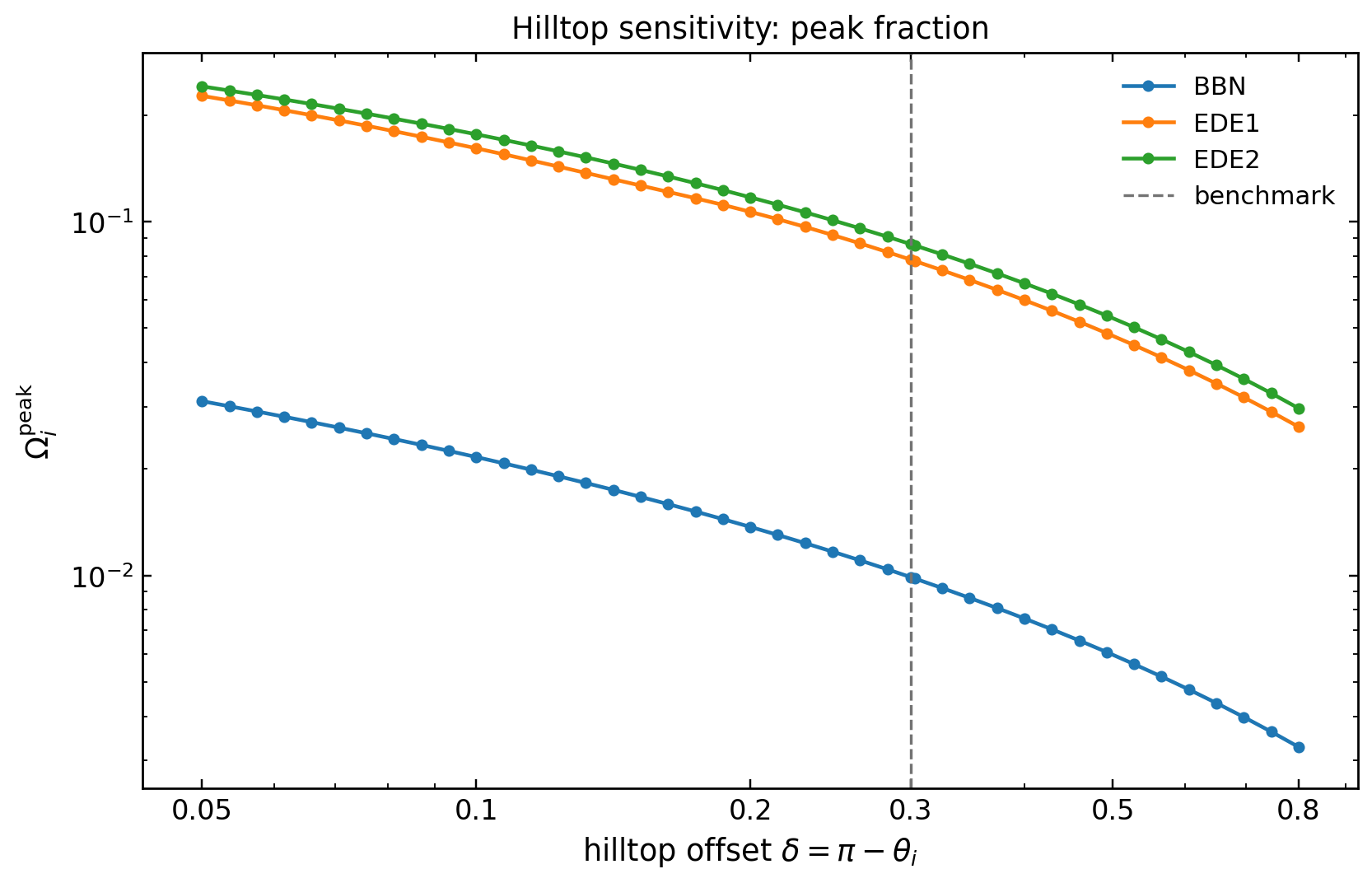}
\caption{Transient peak fractions as functions of the common
initial offset. The three transient initial angles vary
together while all potential parameters and the late-time
initial angle remain fixed. Each marker corresponds to a
four-field background solution; the lines connect the
calculated points. Smaller offsets delay rolling and
increase the fractional-density maxima. The dashed line
marks the benchmark offset \(\delta=0.30\).}
\label{fig:hilltop_omega}
\end{figure}

\begin{figure}[t]
\centering
\includegraphics[width=0.90\textwidth]
{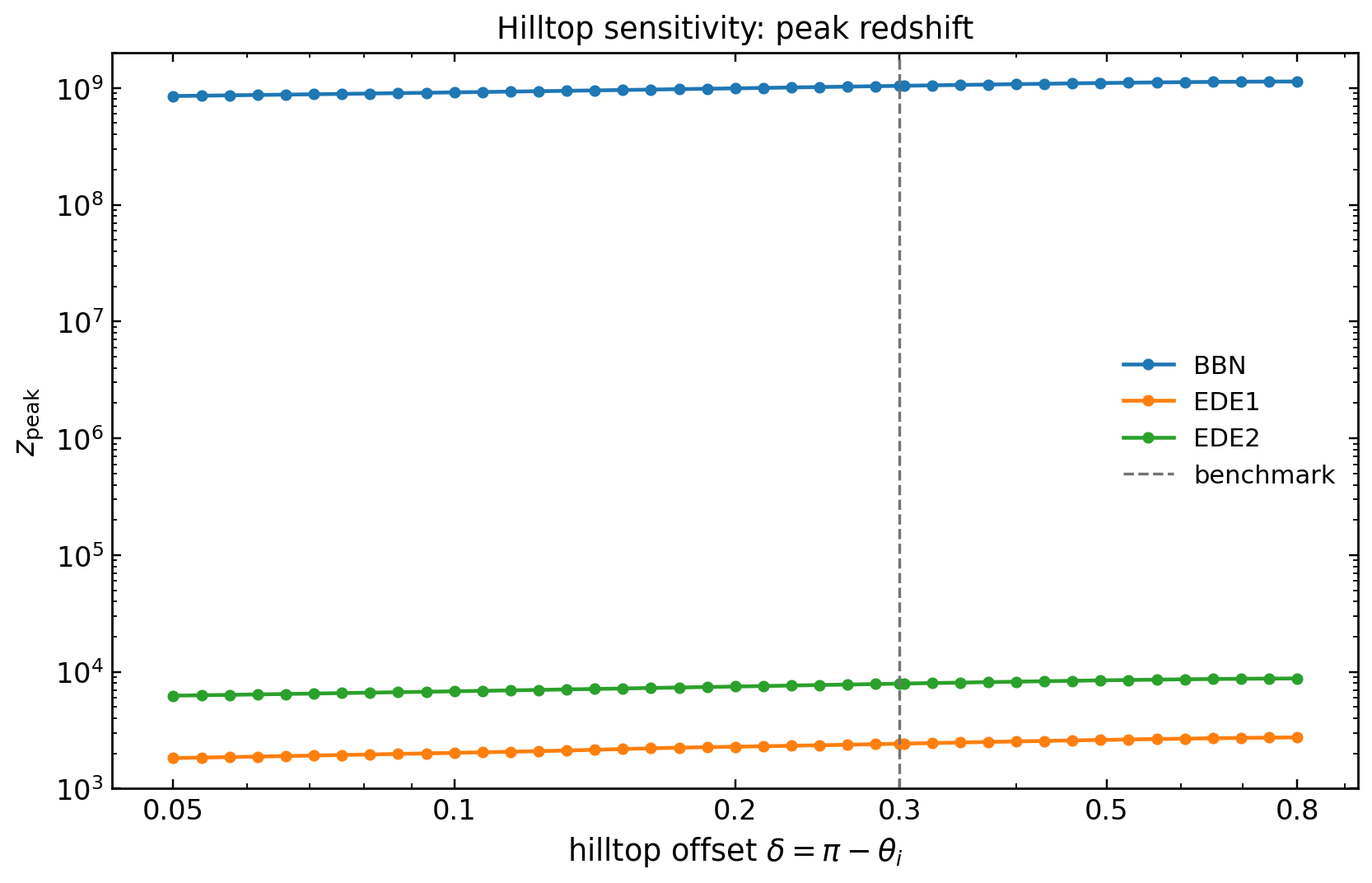}
\caption{Redshifts of the fractional-density maxima in
the same scan as figure~\ref{fig:hilltop_omega}. Larger
offsets produce earlier peaks. The hilltop curvature
scales remain fixed, so the shifts arise from changes
in the rolling delay and the shared expansion history.
The dashed line marks the benchmark offset
\(\delta=0.30\).}
\label{fig:hilltop_z}
\end{figure}

Figure~\ref{fig:hilltop_omega} shows that the peak fractions
decrease monotonically as the common offset increases.
Table~\ref{tab:hilltop_summary} lists the endpoints and
the benchmark. The corresponding peak redshifts are shown
in figure~\ref{fig:hilltop_z}. Larger offsets bring the
peaks to higher redshifts, but the fractional changes in
their epochs are smaller than those in their amplitudes.

\begin{table}[t]
\centering
\small
\setlength{\tabcolsep}{4pt}
\caption{Peak redshifts and peak energy fractions at the benchmark offset and the two endpoints of the scan. Each pair refers to an individual field's maximum fractional density; the three maxima in each row occur at different epochs.}
\label{tab:hilltop_summary}
\begin{tabular}{ccccccc}
\toprule
& \multicolumn{2}{c}{BBN}
& \multicolumn{2}{c}{EDE1}
& \multicolumn{2}{c}{EDE2} \\
\(\delta\)
& \(z_{\rm peak}\) & \(\Omega^{\rm peak}\)
& \(z_{\rm peak}\) & \(\Omega^{\rm peak}\)
& \(z_{\rm peak}\) & \(\Omega^{\rm peak}\) \\
\midrule
0.05
& \(8.52\times10^8\) & 0.0311
& \(1.83\times10^3\) & 0.227
& \(6.25\times10^3\) & 0.241 \\
0.30
& \(1.04\times10^9\) & 0.00990
& \(2.43\times10^3\) & 0.0781
& \(7.93\times10^3\) & 0.0864 \\
0.80
& \(1.13\times10^9\) & 0.00328
& \(2.75\times10^3\) & 0.0263
& \(8.78\times10^3\) & 0.0297 \\
\bottomrule
\end{tabular}
\end{table}

The responses include the gravitational influence of each
field on the others through the shared expansion rate.
The common-offset scan measures their joint dependence
on the initial conditions; the separate field responses
relevant to isocurvature are discussed in
section~\ref{subsec:isocurvature}.

Across the sampled range, the peak fractions vary by
factors of approximately \(8\)--\(10\), whereas the peak
redshifts vary by factors of approximately \(1.3\)--\(1.5\).
The sequence of the BBN, heavier EDE, and lighter EDE
peaks is preserved. Thus, for these fixed potentials,
the initial displacement affects the strength of each
transient more strongly than its characteristic epoch.
% ============================================================
\section{Aligned inflation: field range and initial conditions}
\label{app:inflation}

We derive the effective field range of the aligned sector
introduced in section~\ref{subsec:inflation_connection}
and examine how the adopted inflationary and reheating
scales determine the angular fluctuations and Hubble damping
of the four post-inflationary fields.
\subsection{Alignment and the effective period}
For two canonical microscopic fields \(\varphi_1\) and \(\varphi_2\), we consider
\begin{align}
V_{\rm inf}
&=\Lambda_A^4\left[
1-\cos\left(
p_1\frac{\varphi_1}{f_{\varphi_1}}
+p_2\frac{\varphi_2}{f_{\varphi_2}}
\right)\right]
\nonumber\\
&\quad+\Lambda_B^4\left[
1-\cos\left(
q_1\frac{\varphi_1}{f_{\varphi_1}}
+q_2\frac{\varphi_2}{f_{\varphi_2}}
\right)\right]
+\delta V_{\rm flat}.
\label{eq:aligned_potential}
\end{align}
Here \(f_{\varphi_1}\) and \(f_{\varphi_2}\) are the microscopic
decay constants, \(\boldsymbol p=(p_1,p_2)\) and
\(\boldsymbol q=(q_1,q_2)\) are integer charge vectors, and
\(\Lambda_A\) and \(\Lambda_B\) set the potential scales.
The term \(\delta V_{\rm flat}\) denotes a possible deformation of the potential along the light direction. The evolution of the aligned field during inflation can be
calculated once the potential and its energy scales are
specified.

We assume that the first cosine stabilizes one field
combination near its minimum, with a mass much larger
than the Hubble rate during inflation. In the canonically
normalized field plane, we define
\begin{equation}
\mathcal N_p\equiv
\sqrt{
\frac{p_1^2}{f_{\varphi_1}^2}
+\frac{p_2^2}{f_{\varphi_2}^2}
},
\qquad
\boldsymbol e_L\equiv
\frac{1}{\mathcal N_p}
\left(
\frac{p_2}{f_{\varphi_2}},
-\frac{p_1}{f_{\varphi_1}}
\right).
\label{eq:aligned_light_direction}
\end{equation}
The unit vector \(\boldsymbol e_L\) is orthogonal to the
gradient of the first cosine's phase. A canonical displacement \(\ell\) along this direction changes the second phase by
\begin{equation}
\ell\left[
\frac{q_1}{f_{\varphi_1}}e_{L,1}
+\frac{q_2}{f_{\varphi_2}}e_{L,2}
\right]
=
\ell\,
\frac{p_2q_1-p_1q_2}
{\sqrt{
p_2^2f_{\varphi_1}^2+p_1^2f_{\varphi_2}^2
}},
\label{eq:aligned_projection}
\end{equation}
where \(e_{L,1}\) and \(e_{L,2}\) are the components of
\(\boldsymbol e_L\). The second cosine therefore induces a
period \(2\pi f_{\rm eff}\), with
\begin{equation}
f_{\rm eff}
=
\frac{
\sqrt{p_2^2f_{\varphi_1}^2+p_1^2f_{\varphi_2}^2}
}{
|p_2q_1-p_1q_2|
}.
\label{eq:feff}
\end{equation}
A small nonzero determinant of the charge vectors enhances
the effective period relative to the microscopic field ranges
\cite{Kim:2004rp,Choi:2014rja}.

For the choice
\begin{equation}
\boldsymbol p=(10,11),
\qquad
\boldsymbol q=(9,10),
\qquad
f_{\varphi_1}=f_{\varphi_2}=0.5\Mpl,
\end{equation}
the determinant has unit magnitude, giving
\(f_{\rm eff}\simeq7.43\Mpl\).
Alignment therefore produces a super-Planckian effective
decay constant along the light field direction, while
the microscopic axion decay constants remain sub-Planckian.

We assume that the heavy field stays near its minimum as
the light field rolls. The potential scales and
\(\delta V_{\rm flat}\) must preserve this heavy--light
mass hierarchy. If the heavy field is excited, for example
by a sharp turn in the trajectory, both fields must be
evolved. 

The alignment example illustrates the enhancement of the
effective field range. With the heavy direction fixed and
\(\delta V_{\rm flat}=0\), the light-field potential reduces
to an ordinary cosine. For \(f_{\rm eff}\simeq7.43\Mpl\)
and 50--60 e-folds between horizon exit of the reference
mode and the end of inflation, this potential predicts
\(r\simeq0.073\)--\(0.097\), above the published
BICEP/Keck 2018 bound
\cite{BICEP:2021xfz}. The value \(r=0.01\) used below
serves as a separate fiducial normalization for the
inflationary-scale and field-fluctuation estimates.
Obtaining this value from the aligned sector would
require a specified deformation \(\delta V_{\rm flat}\)
and the corresponding inflationary trajectory.

\subsection{Inflationary energy scale}

We denote the curvature power-spectrum amplitude by \(A_s\)
and the tensor-to-scalar ratio by \(r\), both evaluated at
a reference comoving wavenumber \(k_*\). The potential
when this mode exits the Hubble radius, \(k_*=aH\), is
denoted by \(V_*\). For slow roll along the light direction,
with negligible contributions from other fields to the
curvature perturbation, the potential slow-roll parameter is
\begin{equation}
\epsilon_*
=
\frac{\Mpl^2}{2}
\left(\frac{V_{,\ell}}{V}\right)_*^2,
\end{equation}
where \(V\) is evaluated along the effective light-field
trajectory. At leading order in the slow-roll approximation,
\begin{equation}
A_s\simeq
\frac{V_*}{24\pi^2\Mpl^4\epsilon_*},
\qquad
r\simeq16\epsilon_*.
\label{eq:slowroll_normalization}
\end{equation}
Using \(3\Mpl^2H_{\rm inf}^2\simeq V_*\), we obtain
\begin{equation}
V_*^{1/4}\simeq
\left(\frac{3\pi^2}{2}A_s r\right)^{1/4}\Mpl,
\qquad
H_{\rm inf}\simeq
\pi\Mpl\sqrt{\frac{A_s r}{2}}.
\label{eq:inflation_scales}
\end{equation}

For the inflationary-scale estimates, we adopt
\(A_s=2.10\times10^{-9}\) and the fiducial tensor ratio
\(r=0.01\) at \(k_*=0.05\,\mathrm{Mpc}^{-1}\).
The corresponding energy and Hubble scales are
\begin{equation}
V_*^{1/4}\simeq1.02\times10^{16}\,\GeV,
\qquad
H_{\rm inf}\simeq2.48\times10^{13}\,\GeV.
\label{eq:inflation_numbers}
\end{equation}
The adopted tensor ratio lies below the published BICEP/Keck
2018 limit \(r_{0.05}<0.036\) at 95\% confidence
\cite{BICEP:2021xfz,Planck:2018jri}. The angular fluctuations have amplitude
\(H_{\rm inf}/(2\pi f_i)\). Using \(r=0.01\) and the
benchmark decay constants, we obtain approximately
\(4.1\times10^{-5}\) for the BBN field,
\(1.3\times10^{-5}\) for each EDE field, and
\(1.1\times10^{-6}\) for the late-time field.
These provide the initial field fluctuations considered in
section~\ref{subsec:isocurvature}.

\subsection{Reheating and the lower-curvature fields}

We take reheating to end at a fiducial temperature
\(T_{\rm RH}=10^9\,\GeV\), after which radiation dominates
the expansion. For an effective number of relativistic energy
degrees of freedom \(g_{*,{\rm RH}}=106.75\), the radiation
density is
\(\rho_{r,{\rm RH}}=(\pi^2/30)g_{*,{\rm RH}}T_{\rm RH}^4\).
From the Friedmann equation, we obtain
\begin{equation}
H_{\rm RH}
=
\left(\frac{\pi^2g_{*,{\rm RH}}}{90}\right)^{1/2}
\frac{T_{\rm RH}^2}{\Mpl}
\simeq1.405\,\GeV.
\label{eq:reheating_H}
\end{equation}
Comparing this rate with the characteristic curvature scales
indicates whether Hubble damping suppresses the fields' motion at the onset of radiation domination. The corresponding ratios for the four post-inflationary fields are listed in table~\ref{tab:reheating}.

\begin{table}[t]
\centering
\caption{Ratios of the Hubble rate to the characteristic field scales at \(T_{\rm RH}=10^9\,\GeV\). We use the hilltop curvature scales for the three transients and the vacuum mass for the late-time field. All four ratios are much larger than unity, placing the fields in the strongly overdamped regime for the adopted potentials.}
\label{tab:reheating}
\begin{tabular}{lc}
\toprule
Field &
\(H_{\rm RH}/m_{\rm eff}\) or \(H_{\rm RH}/m_{\rm DE}\) \\
\midrule
BBN       & \(1.24\times10^{25}\) \\
EDE1      & \(1.62\times10^{36}\) \\
EDE2      & \(1.87\times10^{35}\) \\
Late time & \(9.63\times10^{41}\) \\
\bottomrule
\end{tabular}
\end{table}

The large Hubble-to-curvature ratios strongly suppress
potential-driven rolling during reheating. When the
potential gradient is negligible, the scalar equation
becomes \(\ddot\phi_i+3H\dot\phi_i\simeq0\).
An initial velocity then decreases as
\(\dot\phi_i\propto a^{-3}\), and its kinetic energy
density dilutes as \(a^{-6}\). Hubble damping can therefore
leave the fields displaced from their minima with
negligible velocities. If the adopted misalignment angles
are preserved through reheating, as discussed in
section~\ref{subsec:isocurvature}, the subsequent four-field
evolution can begin from the nearly frozen initial
conditions used here.

% ============================================================
\section{Numerical evolution and convergence checks}
\label{app:numerics}
% ============================================================

In this section we describe the numerical evolution of the four-field background, from the initially frozen configuration to the present epoch. The equations are written in terms of the logarithmic scale factor, allowing the BBN, EDE, and late-time contributions to be followed within the same integration. We then present the initial conditions and convergence checks used to determine
the transient peaks, residual densities, and late-time
equation of state.

\subsection{Evolution equations}

We use \(N\equiv\ln a\), with a prime denoting
\(\dd/\dd N\). Since \(\dot N=H\), the time derivatives
satisfy
\begin{equation}
\frac{\dd}{\dd t}=H\frac{\dd}{\dd N},
\qquad
\dot\phi_i=H\phi_i',
\qquad
\ddot\phi_i=H^2\phi_i''+HH'\phi_i'.
\label{eq:time_conversion}
\end{equation}
Substituting these expressions into the Klein--Gordon
equation and dividing by \(H^2\) gives the following result
\begin{equation}
\phi_i''+\left(3+\frac{H'}{H}\right)\phi_i'
+\frac{V_{i,\phi}}{H^2}=0.
\label{eq:kg_N}
\end{equation}
The additional coefficient \(H'/H\) accounts for the changing
relation between cosmic time and \(N\).

In the same variables, the scalar density and pressure are
\begin{equation}
\rho_{\phi_i}=\frac12H^2\phi_i'^2+V_i,
\qquad
P_{\phi_i}=\frac12H^2\phi_i'^2-V_i.
\label{eq:rhoP_N}
\end{equation}
In these variables, the scalar kinetic energies contain a
factor of \(H^2\). We collect all terms proportional to
\(H^2\) on the left-hand side of the Friedmann equation:
\begin{equation*}
H^2\left(3\Mpl^2-\frac12\sum_i\phi_i'^2\right)
=\rho_{r0}e^{-4N}+\rho_{m0}e^{-3N}+\sum_iV_i.
\end{equation*}
Therefore we get,
\begin{equation}
H^2=
\frac{\rho_{r0}e^{-4N}+\rho_{m0}e^{-3N}+\sum_iV_i}
{3\Mpl^2-\frac12\sum_i\phi_i'^2}.
\label{eq:H_N}
\end{equation}
This expression determines \(H\) algebraically from the
field values, their derivatives, and the matter and
radiation densities at each integration step. The
remaining coefficient in eq.~\eqref{eq:kg_N} follows
from \(H'/H=\dot H/H^2\):
\begin{equation}
\frac{H'}{H}=-
\frac{\frac43\rho_r+\rho_m+H^2\sum_i\phi_i'^2}
{2\Mpl^2H^2}.
\label{eq:Hprime_N}
\end{equation}
Both the potential and kinetic energies of the four
fields therefore enter the expansion rate governing
their motion.

For the numerical solution, we express the system in
dimensionless form:
\begin{equation}
\begin{aligned}
u_i&\equiv\Theta_i',
\qquad
F_i\equiv\frac{f_i}{\Mpl},
\qquad
E\equiv\frac{H}{H_0},
\\
\alpha_i&\equiv\frac{\Lambda_i^4}{3\Mpl^2H_0^2},
\qquad
K_i\equiv\frac{F_i^2u_i^2}{6}.
\end{aligned}
\label{eq:dimensionless_variables}
\end{equation}
Since \(\phi_i'=f_i u_i\), the quantity
\(K_i=\dot\phi_i^2/(2\rho_{\rm tot})\) is the fraction
of the total density carried by the kinetic energy of
the \(i\)th field. Equation~\eqref{eq:H_N} becomes
\begin{equation}
E^2=
\frac{\Omega_{r0}e^{-4N}+\Omega_{m0}e^{-3N}
+\sum_i\alpha_i(1-\cos\Theta_i)^{n_i}}
{1-\sum_iK_i}.
\label{eq:dimensionless_friedmann}
\end{equation}
The instantaneous fluid fractions are
\(\Omega_r=\Omega_{r0}e^{-4N}/E^2\) and
\(\Omega_m=\Omega_{m0}e^{-3N}/E^2\), so
eq.~\eqref{eq:Hprime_N} reduces to
\begin{equation}
\frac{H'}{H}
=-2\Omega_r-\frac32\Omega_m-3\sum_iK_i.
\label{eq:dimensionless_hprime}
\end{equation}

To obtain the angular equations, we substitute
\(\phi_i=f_i\Theta_i\) into eq.~\eqref{eq:kg_N}
and differentiate the periodic potential. Its derivative
with respect to \(\Theta_i\) is
\(n_i\Lambda_i^4(1-\cos\Theta_i)^{n_i-1}\sin\Theta_i\).
Therefore, the resulting first-order system is
\begin{align}
\Theta_i'&=u_i,
\nonumber\\
u_i'&=-\left(3+\frac{H'}{H}\right)u_i
-\frac{3\alpha_i n_i}{F_i^2E^2}
(1-\cos\Theta_i)^{n_i-1}\sin\Theta_i.
\label{eq:dimensionless_kg}
\end{align}
These eight equations are evolved together, with
\(E\) and \(H'/H\) evaluated from the combined field
and fluid densities.

The scalar fractions and equations of state are
reconstructed by dividing the density and pressure
in eq.~\eqref{eq:rhoP_N} by \(3\Mpl^2H^2\):
\begin{equation}
\begin{aligned}
\Omega_i
&=K_i+\frac{\alpha_i(1-\cos\Theta_i)^{n_i}}{E^2},
\\
w_i
&=\frac{K_i-\alpha_i(1-\cos\Theta_i)^{n_i}/E^2}
{K_i+\alpha_i(1-\cos\Theta_i)^{n_i}/E^2}.
\end{aligned}
\label{eq:dimensionless_outputs}
\end{equation}
The small residual densities require accurate evaluation
of the potential near its minima. We therefore calculate
\(1-\cos\Theta_i\) as \(2\sin^2(\Theta_i/2)\), avoiding
the loss of precision from subtracting nearly equal
numbers. We also check that \(1-\sum_iK_i>0\), as
required by eq.~\eqref{eq:dimensionless_friedmann}
for the positive-density components considered here.

Scalar energy conservation follows from differentiating
\(\rho_{\phi_i}\) with respect to \(N\) and substituting
eq.~\eqref{eq:kg_N}:
\begin{equation}
\rho_{\phi_i}'=-3H^2\phi_i'^2
=-3(\rho_{\phi_i}+P_{\phi_i}).
\label{eq:scalar_conservation_N}
\end{equation}
Since the total density obeys
\(\rho_{\rm tot}'=-3(1+w_{\rm tot})\rho_{\rm tot}\),
the fractional density evolves according to
\(\Omega_i'=3\Omega_i(w_{\rm tot}-w_i)\), as in
eq.~\eqref{eq:fraction_peak_condition}.
A positive-to-negative crossing of \(w_{\rm tot}-w_i\)
marks a local maximum of \(\Omega_i\). We use these
crossings to locate the transient peaks numerically.

\subsection{Initial conditions, integration, and convergence}

The integration begins at \(z_{\rm ini}=10^{11}\), with
the initial displacements specified in
table~\ref{tab:benchmark} and vanishing field velocities:
\begin{equation}
N_{\rm ini}=-\ln(1+z_{\rm ini}),
\qquad
\Theta_i(N_{\rm ini})=\theta_i,
\qquad
u_i(N_{\rm ini})=0.
\label{eq:initial_conditions}
\end{equation}
The late-time potential scale is adjusted to satisfy
the present-day normalization \(H(N=0)=H_0\), including
the residual densities of the three transient fields.
The value retained in the integration is
\begin{equation}
\Lambda_{\rm DE}
=2.259386173961176\times10^{-3}\,\eV,
\label{eq:lambda_internal}
\end{equation}
The resulting solution satisfies
\(\lvert H(N=0)/H_0-1\rvert<8\times10^{-11}\).
The additional digits specify the input needed to
reproduce this numerical normalization.

The first-order system is integrated to the present epoch,
\(N=0\), with a relative tolerance of \(10^{-10}\) and an
absolute tolerance of \(10^{-13}\) for each dimensionless
variable. The step size is limited to \(\Delta N\leq0.02\).
We retain the full periodic potentials through the initial
roll and subsequent oscillations.

Oscillations can produce secondary fractional-density maxima.
For each field, we select the largest maximum identified
using eq.~\eqref{eq:fraction_peak_condition} and check
the peak extraction by doubling the search-grid resolution.
We test convergence by repeating the integration with relative
and absolute tolerances of \(10^{-9}\) and \(10^{-12}\),
respectively, and a maximum step \(\Delta N=0.04\).
The sampled scalar fractions differ by at most
\(4.9\times10^{-6}\) in relative terms, with differences
below \(1.5\times10^{-6}\) at the tabulated epochs.
The peak fractions and redshifts change by less than
\(8\times10^{-11}\).

To check the dependence on the initial conditions,
we repeat the evolution from \(z=10^{12}\), again with
vanishing initial velocities, and integrate to \(z=100\).
Relative to the run starting at \(z=10^{11}\), the
transient peak fractions and redshifts change by less
than \(1.4\times10^{-7}\).

\subsection{Hilltop scan and late-time fit}

For the common-offset scan in appendix~\ref{app:hilltop},
we evolve all four fields from \(z_{\rm ini}=10^{11}\)
to \(z=100\). The potential parameters and late-time
initial angle remain fixed, including \(\Lambda_{\rm DE}\)
in eq.~\eqref{eq:lambda_internal}. The benchmark \(H_0\)
remains the reference scale in the dimensionless equations.

Repeating each point with both integration settings and
doubling the peak-search grid changes the peak fractions
and redshifts by at most \(3.8\times10^{-9}\) in relative
terms. Earlier-start runs from \(z=10^{12}\) at
\(\delta=0.05,0.30,0.80\) change these quantities by
less than \(1.6\times10^{-7}\).

The CPL coefficients in eq.~\eqref{eq:w0wa_result}
are obtained by unweighted least squares using 10,001
values of \(w_{\rm DE}(a)\), uniformly spaced in
\(\ln a\) over \(1/3\leq a\leq1\). The largest absolute
fit residual is \(4.85\times10^{-3}\).

\bibliographystyle{JHEP}
\bibliography{references}

\end{document}